\documentclass[pra,twocolumn,10pt,superscriptaddress,aps]{revtex4-1}
\usepackage{amsmath}
\usepackage{amssymb}
\usepackage[latin1]{inputenc}
\usepackage{textcomp}
\usepackage{graphicx}
\usepackage{wasysym}
\usepackage{physics}
\usepackage{stackrel}
\usepackage{multirow}
\usepackage{braket,mleftright}
\usepackage{empheq}
\usepackage{subfigure}
\usepackage{blkarray}
\usepackage{bm}
\usepackage{bbold}
\usepackage{color}
\usepackage[unicode=true,
 bookmarks=false,
 breaklinks=false,pdfborder={0 0 1},backref=false,colorlinks=false]
 {hyperref} 
\usepackage{breakurl}
\usepackage{dsfont}
\usepackage{soul}

\makeatletter

\usepackage{amsfonts}
\usepackage{amsthm}
\usepackage{graphics}
\usepackage{latexsym}
\usepackage{subfigure}
\usepackage{braket}
\usepackage{float}
\usepackage{palatino}
\usepackage{bbm}
\usepackage{calrsfs}
\usepackage[inline]{enumitem}
\usepackage[mathscr]{eucal}
\usepackage{silence}
\newtheorem{theorem}{Theorem}

\newtheorem{definition}[theorem]{Definition}

\newcommand{\p}[1]{\textup{p}\left( #1 \right)}
\newcommand{\projector}[1]{\ketbra{#1}}
\newcommand{\floor}[1]{\left\lfloor #1 \right\rfloor}

\makeatother

\begin{document}

\title{Certification of continuous-variable gates using average channel-fidelity witnesses}

\author{Renato M. S. Farias}
\email{renato.msf@gmail.com}
\affiliation{Instituto de F\'{i}sica, Universidade Federal do Rio de Janeiro, P. O. Box 68528, Rio de Janeiro, Rio de Janeiro 21941-972, Brazil}

\author{Leandro Aolita}
\affiliation{Instituto de F\'{i}sica, Universidade Federal do Rio de Janeiro, P. O. Box 68528, Rio de Janeiro, Rio de Janeiro 21941-972, Brazil}
\affiliation{Quantum Research Centre, Technology Innovation Institute, Abu Dhabi, UAE}

\begin{abstract}
We introduce witnesses for the average channel fidelity between a known target gate and an arbitrary unknown channel, for continuous-variable (CV) systems. These are observables whose expectation value yields a tight lower bound to the average channel fidelity in question, thus constituting a practical tool for certification of experimental CV gates. Our framework applies to a broad class of target gates. Here, we focus on three specific types of targets: multi-mode Gaussian unitary channels, single-mode coherent state amplifiers, and the single-mode (non-Gaussian) cubic phase gate, which is a crucial ingredient for CV universal quantum computation. Our witnesses are experimentally-friendly as they rely exclusively on Gaussian measurements, even for the non-Gaussian-target case. Moreover, in all three cases, they can be measured efficiently in the estimation error $\epsilon$ and failure probability $\Delta$, as well as in the number of modes $m$ for the Gaussian-target case.
To end up with, our approach for the Gaussian-target case relies on an improved measurement scheme for Gaussian state-fidelity witnesses, which is polynomially and exponentially more efficient in $m$ and $\Delta$, respectively, than previous schemes. The latter constitutes an interesting byproduct result on its own.
Our findings are relevant to the experimental validation of many-body quantum technologies.
\end{abstract}

\maketitle

%%%%%%%%%%%%%%%%%%%%%%%%%%%%%%%%%%%%%%%%%%%%%%%%%%%%%%%%%%%%%%%%%%%%%%%%%%%%%%%%%%%%%%%%%%%%%%%
\section{Introduction}
\label{SectionI}
%%%%%%%%%%%%%%%%%%%%%%%%%%%%%%%%%%%%%%%%%%%%%%%%%%%%%%%%%%%%%%%%%%%%%%%%%%%%%%%%%%%%%%%%%%%%%%%

The validation of experimentally implemented gates is one of the major current bottlenecks in the development of many-body quantum simulators as well as universal quantum computers.
In principle, an arbitrary unknown channel can be fully characterized with quantum process tomography \cite{Chuang1997, Poyatos1997, Mohseni2008}. However, this is in practice not a feasible option for multi-partite or high-dimensional single-partite systems \cite{Bendersky2013, Namiki2016}: It requires the tomographic reconstruction of the channel's output state either for a large number of different input states  \cite{Lopez2010, Schmiegelow2011} or for the input system in a maximally entangled state with an ancilla \cite{Altepeter2003, Bai2018}. In either case, quantum state tomography of a single output state requires already a number of measurements that scales very unfavorably with the system dimension. This is specially problematic for continuous-variable (CV) systems \cite{Lvovsky2009}.

For the case of state preparations, several techniques have been put forward for validation or benchmarking in order to avoid full quantum state tomography \cite{Namiki2008, Cramer2010, Gross2010, Toth2010, Flammia2011, Silva2011, Moroder2012, Flammia2012, Reich2013, Aolita2015, Bai2018, Gluza2018, Liu2018}. However, in contrast, certification of quantum gates is a much less explored field. For discrete-variable systems, a remarkable technique for characterizing  average gate-error rates in circuits with variable (random) components is randomized benchmarking \cite{Emerson2005, Knill2008, Dankert2009, Magesen2011, Cross2016}. However, this does not allow one to certify a single target circuit with fixed components. Moreover, for CV systems the problem is less understood \cite{Lvovsky2009, Aolita2015, Bai2018, Roth2018, Sharma2018}.

Here, we derive experimentally-friendly observables on CV systems whose expectation value yields a tight lower bound to the \textit{average channel fidelity} \cite{Braunstein1999, Nielsen2002, Pirandola2015} between a known target gate and an arbitrary unknown channel. We refer to these as \textit{average channel-fidelity witnesses}, in analogy to state-fidelity witnesses \cite{Aolita2015, Gluza2018, Liu2018}. The method extends to a wide spectrum of target gates. In particular, here, we explicitly present efficiently-measurable witnesses for  three classes of targets: multi-mode \textit{Gaussian unitary channels} \cite{Ferraro2005, Cerf-book2007, Weedbrook2011}, single-mode \textit{coherent state amplifiers} \cite{Hammerer2005, Namiki2008, Pooser2009, Ferreyrol2010, Zavatta2010, Chiribella2013}, and the single-mode (non-Gaussian) \emph{cubic phase gate} \cite{Gottesman2001, Gu2009, Weedbrook2011}.
Importantly, the latter is highly relevant for computations, since, together with Gaussian channels, it is enough for CV universal quantum computing \cite{Lloyd1999}.
Our witnesses are highly experimentally-friendly in that they can be measured by probing the channel with simple Gaussian states (coherent states) as inputs and making Gaussian measurements (homodyne detection) \cite{Ferraro2005, Cerf-book2007, Weedbrook2011} at the  output, even for the non-Gaussian target gates considered. 
Moreover, in all three cases, the estimation of their expectation value is efficient in all relevant parameters: its \textit{sample complexity} (i.e. number of experimental runs required) scales polynomially in the inverse estimation error $1/\epsilon$ and logarithmically in the inverse failure probability $1/\Delta$, as well as polynomially in the number of modes $m$ for the Gaussian-target case. Such scaling in $\Delta$ represents an  exponential improvement with respect to previous estimation methods \cite{Aolita2015}, and is possible thanks to an estimation protocol known as \textit{median-of-means} \cite{Nemirovsky1983, Jerrum1986}.
Furthermore, our measurement scheme for witnesses of Gaussian target gates exploits state-fidelity witnesses for Gaussian target states \cite{Aolita2015}. To measure the latter, we use an enhanced method \cite{Gluza2018} based on importance sampling [see, \emph{e.g.}, Refs. \cite{Owen2000, Tokdar2010}], which (apart from the already mentioned improvement on $\Delta$) is polynomially more efficient in $m$ than previous methods \cite{Aolita2015}. This is interesting in its own right beyond the scope of channel certification.

The paper is organized as follows: in Sec. \ref{SectionII} we introduce the notation and background. In Sec. \ref{SectionIII} we state our main theorems. The estimation protocols are detailed in Sec. \ref{SectionIV}. To end up with, our conclusions are left for Sec. \ref{SectionVI}.

%%%%%%%%%%%%%%%%%%%%%%%%%%%%%%%%%%%%%%%%%%%%%%%%%%%%%%%%%%%%%%%%%%%%%%%%%%%%%%%%%%%%%%%%%%%%%%%
\section{Background and Notation}
\label{SectionII}
%%%%%%%%%%%%%%%%%%%%%%%%%%%%%%%%%%%%%%%%%%%%%%%%%%%%%%%%%%%%%%%%%%%%%%%%%%%%%%%%%%%%%%%%%%%%%%%

For an $m$-mode CV state described by a density matrix $\varrho$, each mode can be described by a pair of bosonic field operators $\{a_{k}^{\dagger}, \, a_{k}\}_{k \in [m]}$, where $a_{k}^{\dagger} \, (a_{k})$ is the $k$-th \textit{creation} (\textit{annihilation}) operator, and $[m] = \{1, 2, \cdots, m\}$. The field operators satisfy the usual commutation relations $[a_{k}, a_{l}^{\dagger}] = \delta_{k l}$, where $\delta$ is the \textit{Kronecker delta}. The multi-mode number operator, $n$, is defined as the sum of all single-mode number operators, \textit{i.e.} $n \coloneqq \sum_{k=1}^{m} a_{k}^{\dagger}a_{k}$. Operators $q_{k} = (a_{k} + a_{k}^{\dagger})/2$ and $p_{k} = (a_{k} - a_{k}^{\dagger})/2i$ are, respectively, the $k$-th mode's position- and momentum-like quadrature field operators in the \textit{quantum optical convention}, \textit{i.e.} $[q_{k}, p_{l}] = i \delta_{k,l} / 2$. The quadrature field operators can be arranged in a \textit{quadrature vector}, $\mathbf{r}$, such as
\begin{equation}
\mathbf{r} \coloneqq \left( \, q_{1}, \, p_{1}, \, q_{2}, \, p_{2}, \cdots, \, q_{m}, \, p_{m} \, \right)^{T}. \label{eq:quadratures}
\end{equation}
We can define the \textit{first-moment vector}, $\mathbf{x} \in \mathbb{R}^{2m}$, as
\begin{equation}
\mathbf{x} \coloneqq \braket{\mathbf{r}}_{\varrho} = \tr(\mathbf{r} \, \varrho), \label{eq:firstMoments}
\end{equation}
where $\tr(\cdot)$ is the trace operation. We can define a $2\,m \times 2\,m$ \textit{second-moment matrix} $\bm{\Gamma}$ with elements $\Gamma_{k l}$ composed of bilinear combinations of quadrature operators:
\begin{equation}
\Gamma_{kl} \coloneqq \frac{1}{2} \braket{ r_{k} \, r_{l} + r_{l} \, r_{k} }_{\varrho}. \label{eq:secondMoments}
\end{equation}
The $2\,m \times 2\,m$, symmetric \textit{covariance matrix} $\mathbf{V}$ is defined as
\begin{equation}
\mathbf{V} \coloneqq \bm{\Gamma} - \mathbf{x}\,\mathbf{x}^{T}. \label{eq:covarianceMatrix}
\end{equation}
The combination of $\mathbf{V}$ and $\mathbf{x}$ contain all information necessary to characterize a Gaussian state \cite{Wick1950}. If $\mathbf{V}$ is the covariance matrix of an arbitrary pure Gaussian state, the combination of the Williamson's Theo. \cite{Williamson1936} with the Euler Decomposition \cite{Braunstein2005, Arvind1995} guarantees that there is a symplectic matrix $\mathbf{S} \in Sp(2m, \, \mathbb{R})$ such that
\begin{equation}
\mathbf{V} = \frac{1}{4} \, \mathbf{S} \, \mathbf{S}^{T} = \frac{1}{4} \, \mathbf{O} \, \mathbf{D}^{2} \, \mathbf{O}^{-1} \, , \label{eq:WilliamsonEuler}
\end{equation}
with $\mathbf{S} = \mathbf{O} \, \mathbf{D} \, \mathbf{O'}$, with $\mathbf{O},\mathbf{O'} \in \mathbb{R}^{2m \times 2m}$ being orthogonal matrices and $\mathbf{D} \in \mathbb{R}^{2m \times 2m}$ being a positive diagonal matrix such that
\begin{equation}
\mathbf{D} = \bigoplus_{k=1}^{m} \, S\left(\xi_{k}\right) = \bigoplus_{k=1}^{m} \, \left( \begin{matrix}
e^{-\xi_{k}} & 0 \\
0 & e^{\xi_{k}}
\end{matrix} \right) \, , \label{eq:SingleModeSqueezers}
\end{equation}
where $\left\{ S\left(\xi_{k}\right) \right\}_{k \in [m]}$ is the set of single-mode squeezing matrices, and $\xi_{k} \geq 0$ is the $k$-th \textit{single-mode squeezing parameter}. The matrix $\mathbf{S}$ is said to perform a symplectic diagonalization of $\mathbf{V}$ \cite{Williamson1936, Ferraro2005, Weedbrook2011}, and it is equivalent to a corresponding unitary operation $U$ in Hilbert space. Equations \ref{eq:WilliamsonEuler} and \ref{eq:SingleModeSqueezers} are also valid for Gaussian unitary channels.

The state fidelity $F$ between a pure target state $\varrho_{t}$ and a state preparation $\varrho_{p}$ is
\begin{equation}
F \coloneqq F\left(\varrho_{t}, \, \varrho_{p} \right) \coloneqq \tr(\left[ \sqrt{\varrho_{t}} \, \varrho_{p}^{\dagger} \, \sqrt{\varrho_{t}} \right]^{1/2})^{2} = \tr(\varrho_{t} \, \varrho_{p}), \label{eq:stateFidelity}
\end{equation}
where the last equality holds because $\varrho_{t}$ is a pure state. $F$ is an excellent figure of merit for how good is the preparation of a desired state \cite{Nielsen2002, Uhlmann2000}. Here, we use the \textit{average channel fidelity}. This is well known from, e.g., the field of quantum teleportation, where it is used as a practical figure of merit for how good is a teleportation channel \cite{Braunstein1999, Nielsen2002, Pirandola2015}. 
Moreover, we chose not to certify states using \textit{diamond norm} \cite{Nielsen2002} as a figure of merit because, even though more reliable, diamond norm estimation is much more resource-intensive. 

Consider an ensemble $\Omega \coloneqq \left\{ \p{\psi}, \, \ket{\psi} \right\}_{\ket{\psi} \, \in \, \mathscr{S}}$ composed of a finite set $\mathscr{S}$ of pure input states $\ket{\psi}$ and a \textit{prior probability distribution} $\textup{p}$ over $\mathscr{S}$. For instance, $\mathscr{S}$ may correspond to a finite-precision resolution of some bounded-energy continuous set of coherent states in phase space.
The average channel fidelity $\bar{F}_{\Omega}$ between an arbitrary unknown channel $\mathscr{E}$ and a unitary target gate $\mathscr{U}$ with respect to $\Omega$ is then defined as
\begin{equation}
\bar{F}_{\Omega}\left( \mathscr{U}, \, \mathscr{E} \right) \coloneqq \sum_{\ket{\psi} \in \mathscr{S}} \, \p{\psi} \, F\left( \mathscr{U}\left(\projector{\psi}\right), \, \mathscr{E}\left(\projector{\psi}\right) \right), \label{eq:channelFidelity}
\end{equation}
where $F\left( \mathscr{U}\left(\ket{\psi}\right), \, \mathscr{E}\left(\ket{\psi}\right) \right)$ is the state fidelity of Eq. \eqref{eq:stateFidelity}, with $\mathscr{U}\left(\projector{\psi}\right) = U \, \projector{\psi} \, U^{\dagger}$ and $\mathscr{E}\left(\projector{\psi}\right)$ playing, respectively, the roles of  $\varrho_{t}$ and $\varrho_{p}$. Equation \eqref{eq:channelFidelity} thus represents the average fidelity between the outputs of the ideal target channel and its real implementation over the ensemble under consideration. Operationally, $\bar{F}_{\Omega}$ is obtained by drawing an input state $\ket{\psi}\in\mathscr{S}$ according to $\p{\psi}$, calculating the corresponding output states fidelity, and then averaging. 

Clearly, definition \eqref{eq:channelFidelity} generalizes straightforwardly to the case where $\mathscr{S}$ is a continuous set with uncountably many elements. Here we restrict to the case of finite input-state sets because we have in mind a scenario where the elements in $\mathscr{S}$ are sampled with a classical computer. However, the method can also be applied to infinite input-state sets as long as one counts on a practical mechanism to sample from them.

It is also important to note that certification via average-fidelity witnesses, if sucessful, demonstrates the experimental gate is close to the ideal gate with respect to actions on the input ensemble, independent of which states compose the ensemble.

%%%%%%%%%%%%%%%%%%%%%%%%%%%%%%%%%%%%%%%%%%%%%%%%%%%%%%%%%%%%%%%%%%%%%%%%%%%%%%%%%%%%%%%%%%%%%%%
\section{Average channel-fidelity witnesses}
\label{SectionIII}
%%%%%%%%%%%%%%%%%%%%%%%%%%%%%%%%%%%%%%%%%%%%%%%%%%%%%%%%%%%%%%%%%%%%%%%%%%%%%%%%%%%%%%%%%%%%%%%
Here we state our main theorems: namely, the sample complexities of estimating $\bar{F}_{\Omega}$ for multi-mode Gaussian unitary target channels as well as single-mode coherent state amplifiers and cubic phase gates. The core of our general procedure to estimate $\bar{F}_{\Omega}$ consists of sampling an input $\ket{\psi}$ from $\mathscr{S}$ according to the prior distribution $\p{\psi}$, and, then, for each $\ket{\psi}$ drawn, estimating a lower bound to $F\left( \mathscr{U}\left(\ket{\psi}\right), \, \mathscr{E}\left(\ket{\psi}\right) \right)$ by measuring a state-fidelity witness for $\mathscr{U}\left(\ket{\psi}\right)$ as target state on the output. For these reasons, before the average channel-fidelity witnesses we present an improved measurement scheme for the state-fidelity witnesses for multi-mode Gaussian target states originally derived in \cite{Aolita2015}.

%----------------------------------------------------------------------------------------------
\subsection{Improved estimation of state-fidelity witness for pure multi-mode Gaussian target states}
\label{SectionIIIA}
%----------------------------------------------------------------------------------------------

Following Ref. \cite{Gluza2018}, we present the generic notion of \textit{fidelity witness} as follows.

\begin{definition}[Fidelity witness]
\label{definitionFidelityWitness}
An observable $W$ is a fidelity witness for $\varrho_{t}$ if, for $\mathscr{W}\left(\varrho_{p}\right) \coloneqq \tr(W \, \varrho_{p})$, it holds that
\begin{enumerate}
\item $\mathscr{W}\left(\varrho_{p}\right) = 1 \, \Leftrightarrow \, \varrho_{p} = \varrho_{t}$; \label{def:property1}
\item $\mathscr{W}\left(\varrho_{p}\right) \leq F\left(\varrho_{t}, \, \varrho_{p}\right), \; \forall \, \varrho_{p} \,$. \label{def:property2}
\end{enumerate}
\end{definition}

Fidelity witnesses for arbitrary pure Gaussian target states
\begin{equation}
\varrho_{t} = U \, \projector{\mathbf{0}} \, U^{\dagger}, \label{eq:targetState}
\end{equation} 
where $\ket{\mathbf{0}} = \bigotimes_{k=1}^{m} \, \ket{0}$ is the $m$-mode vacuum state and $U$ is an arbitrary Gaussian unitary,  were first presented in Ref. \cite{Aolita2015} as
\begin{equation}
W \coloneqq \mathbbm{1} - U \, n \, U^{\dagger}. \label{eq:stateWitnessHilbert}
\end{equation}
Here, $\mathbbm{1}$ is the identity operator on the $m$-mode Hilbert space and $n$ is the multimode number operator defined in Sec. \ref{SectionII}. Therefore, the fidelity lower bound in this case is such that
\begin{equation}
F \geq \mathscr{W}\left(\varrho_{p}\right) \coloneqq \tr(\varrho_{p} \, W) = \tr( \varrho_{p} \, \left(\mathbbm{1} - U \, n \, U^{\dagger} \right) ). \label{eq:stateLowerBoundHilbert}
\end{equation}
$\mathscr{W}\left(\varrho_{p}\right)$ can also be written in terms of first-moment vectors and second-moment matrices \cite{Aolita2015}:
\begin{equation}
\mathscr{W}\left(\varrho_{p}\right) = 1 + \frac{m}{2} - \frac{1}{4} \, \Tr\left[ \mathbf{V}^{-1}_{t} \left( \mathbf{\Gamma}_{p} - \left( 2 \, \mathbf{x}_{p} - \mathbf{x}_{t} \right) \, \mathbf{x}_{t}^{T} \right) \right] \, , \label{eq:fidelityStates}
\end{equation}
where $\Tr(\cdot)$ is the trace operation over $2\,m \times 2\,m$ matrices, and $\mathbf{x}_{p}$ and $\bm{\Gamma}_{p}$ are, respectively, the first-moment vector and second-moment matrix of $\varrho_{p}$. 
We note that, even though Eq. \eqref{eq:fidelityStates} depends on first and second moments of the prepared state, there is not assumption that $\varrho_{p}$ is a Gaussian state (see Ref. \cite{Aolita2015} for details).
 
Next, we characterize the number of measurements required to estimate $\mathscr{W}\left(\varrho_{p}\right)$ up to  statistical error $\epsilon$ and failure probability $\Delta$, i.e. to obtain an estimate $\mathscr{W}^{*}$ of $\mathscr{W}$ such that $\mathbb{P}\left( \left| \mathscr{W}^{*} - \mathscr{W} \right| \geq \epsilon \right) \leq \, \Delta$, where $\mathbb{P}(\cdot)$ denotes probability. Based on physical grounds, we make the following two assumptions on the experimental state: that it is prepared following an identical and independent procedure from run to run (the \emph{i.i.d. assumption}), and that finite-order statistical moments are bounded. In fact, we will only explicitly need to assume that all fourth-order moments are bounded, \textit{i.e.} $\tr(\Gamma_{kl}^{2} \, \varrho_{p}) \leq \Gamma_{\textup{max}} \coloneqq \max_{kl}{\tr(\Gamma_{kl}^{2} \, \varrho_{p})} < \infty$, and that all single-mode mode energies are upper-bounded by the \emph{maximal single-mode energy} $E_{max}^{(p)}$, \textit{i.e.} $\tr((q_{k}^2 + p_{k}^2) \, \varrho_{p}) \leq E_{max}^{(p)} < \infty, \, \forall \, k \in [m]$. The number of measurements required is formally given by the following theorem, proven in App. \ref{SectionVA}.
We use the Bachmann-Landau asymptotic upper-bound notation $\mathscr{O}$.

\begin{theorem}[Sample complexity of certifying pure Gaussian target states]
\label{theorem:States}
Let $\varrho_{t}$ be an arbitrary, known $m$-mode pure Gaussian target state with maximum single-mode squeezing parameter $\xi_{max}^{(t)} \geq 0$ and first-moment vector $\mathbf{x}_{t}$. Let $\varrho_{p}$ describe an arbitrary, unknown i.i.d preparation with maximum single-mode energy $E_{max}^{(p)}$ and fourth-order moments upper-bounded by $\Gamma_{\textup{max}}$. Then, the total number of measurements required to estimate $\mathscr{W}(\varrho_{p})$ up to error at most $\epsilon > 0$ and failure probability at most $\Delta$ scales as 
\begin{equation}
\mathscr{O}\left( \frac{ s_{t}^{4} \, \left[ m^{4} \, \Gamma_{\textup{max}} + m^{3} \, E_{max}^{(p)} \, \left|\left|\mathbf{x}_{t}\right|\right|_{2}^{2} \right] \, \ln\left( 2 / \Delta \right) }{\epsilon^{2}} \right), \label{eq:theoremStates}
\end{equation}
where $s_{t} \coloneqq \exp( \xi_{max}^{(t)})$ and $||\cdot||_{2}$ is the Euclidean vector norm.
\end{theorem}

Theo. \ref{theorem:States} is the basis of the certification method for multi-mode Gaussian target gates in the next subsection. Still, it is itself interesting for state certification, beyond the scope of channel certification. This is due to the fact that the scaling in Eq. \eqref{eq:theoremStates} significantly outperforms the one of previous protocols in two aspects. First, while in our case the sample complexity scales logarithmically with $\Delta^{-1}$, that of Ref. \cite{Aolita2015} scales as $[\ln(1/(1-\Delta))]^{-1}$, which is approximately equal to $\Delta^{-1}$ for $\Delta$ approaching $0$. Thus our scaling is exponentially better in $\Delta^{-1}$ for the most relevant regime. This is due to the fact that, while in Ref. \cite{Aolita2015} estimation errors are assessed with Chebyshev's bound, the proof here uses the median-of-means estimation protocol combined with the Chernoff bound \cite{Chernoff1952} (see App. \ref{SectionVA} for details). Second, the worst-case scaling of the sample complexity of Ref. \cite{Aolita2015} with the number of modes $m$ is $\mathscr{O}(m^{7})$, whereas Eq. \eqref{eq:theoremStates} scales at worst as $\mathscr{O}(m^{4})$. This corresponds to a polynomial improvement in the number of modes. This is possible thanks to an enhanced measurement scheme, based on importance sampling \cite{Owen2000}, similar to that used in Ref.  \cite{Gluza2018} for fermions. The details of this scheme are given in Sec. \ref{SectionIVA}.

%----------------------------------------------------------------------------------------------
\subsection{Average channel-fidelity witness for Gaussian unitary target channels}
\label{SectionIIIB}
%----------------------------------------------------------------------------------------------
Here, the task in hand is to certify an arbitrary quantum Gaussian unitary channel, $\mathscr{U}$. We consider a protocol that consists in probing an experimental channel $\mathscr{E}$ with an ensemble $\Omega = \{ \p{\psi} , \, \ket{\psi}\}_{\ket{\psi} \in \mathscr{S}}\,$, where $\mathscr{S}$ is a set of pure input states $\ket{\psi}$.We can use the fidelity witness in Eq. \eqref{eq:fidelityStates} to derive a general lower-bound for the average channel fidelity between $\mathscr{U}\left(\projector{\psi}\right)$ and $\mathscr{E}\left(\projector{\psi}\right)$, $\bar{F}_{\Omega} \coloneqq \bar{F}_{\Omega}\left(\mathscr{E},\mathscr{U}\right)$, as 
\begin{eqnarray}
\bar{F}_{\Omega} \geq \bar{\mathscr{W}}_{\Omega}\left(\mathscr{E}\right) & \coloneqq & 1 + \frac{m}{2} - \frac{1}{4} \sum_{\ket{\psi} \in \mathscr{S}} \p{\psi} \Tr\left[ \mathbf{V}_{\mathscr{U}}^{-1} \, \mathbf{\Gamma}_{\mathscr{E}} \right] \nonumber \\
& & - \, \frac{1}{4} \sum_{\ket{\psi} \in \mathscr{S}} \, \p{\psi} \Tr\left[ \mathbf{V}_{\mathscr{U}}^{-1} \mathbf{x}_{\mathscr{U}} \, \mathbf{x}_{\mathscr{U}}^{T} \right] \nonumber \\
& &  \nonumber \\
& & + \, \frac{1}{2} \sum_{\ket{\psi} \in \mathscr{S}} \, \p{\psi} \Tr\left[ \mathbf{V}_{\mathscr{U}}^{-1} \, \mathbf{x}_{\mathscr{E}} \, \mathbf{x}_{\mathscr{U}}^{T}\right], \nonumber \\ 
\label{eq:fidelityChannels}
\end{eqnarray}
where $\mathbf{x}_{\mathscr{U}}(\projector{\psi})$ and $\mathbf{V}_{\mathscr{U}}(\projector{\psi})$ are, respectively, the first-moment vector and covariance matrix of target output states $\mathscr{U}(\projector{\psi})$, and $\mathbf{x}_{\mathscr{E}}(\projector{\psi})$ and $\bm{\Gamma}_{\mathscr{E}}(\projector{\psi})$ are, respectively, the first-moment vector and second-moment matrix of experimental output states $\mathscr{E}\left(\projector{\psi}\right)$. For simplicity, dependencies on $\projector{\psi}$ were omitted in Eq. \eqref{eq:fidelityChannels}.

The lower bound $\bar{\mathscr{W}}$ in \eqref{eq:fidelityChannels} depends on the choice of ensemble $\Omega$. We choose $\Omega$ to be
\begin{equation}
\Omega = \left\{ \p{\bm{\alpha}}, \, \ket{\bm{\alpha}} \right\}_{\ket{\bm{\alpha}} \in \mathscr{S}} \, , \label{eq:Omega}
\end{equation}
where $\ket{\bm{\alpha}} \coloneqq \otimes_{k=1}^{m} \, \ket{\alpha_{k}}$ is a $m$-mode coherent state, with $\ket{\alpha_{k}}$ being the $k$-th single-mode coherent state. As in Sec. \ref{SectionIIIA}, the characterization of the number of measurements required to estimate $\bar{\mathscr{W}}$ up to error $\epsilon$ and failure probability $\Delta$ is done via an empirical estimate $\bar{\mathscr{W}}^{*}$ such that $\mathbb{P}\left(\abs{\bar{\mathscr{W}}^{*} - \bar{\mathscr{W}}} \geq \epsilon \right) \leq \Delta$. We again assume \textit{i.i.d.} preparations and bounded moments up to fourth order, \textit{i.e.} $\Gamma_{\textup{max}} \coloneqq \max_{k,l,\,\bm{\ket{\alpha}} \in \mathscr{S}} \tr(\Gamma_{kl}^{2} \, \mathscr{E}(\projector{\bm{\alpha}})) < \infty$. The theorem below, proven in App. \ref{SectionVB}, summarizes our results.

\begin{theorem}[Sample complexity of certifying Gaussian unitary channels]
\label{theorem:Channels}
Let $\mathscr{U}$ be an arbitrary, known target Gaussian unitary with maximum single-mode squeezing parameter $\xi_{max}^{\mathscr{U}} \geq 0$ and maximum single-mode energy $E_{max}^{\mathscr{U}}$. Let $\mathscr{E}$ be the experimentally implemented channel, with maximum single-mode energy $E_{max}^{\mathscr{E}}$ and fourth-order moments upper-bounded by $\Gamma_{\textup{max}}$. In addition, let $\Omega$, defined in Eq. \eqref{eq:Omega}, be the probe ensemble used. Then, the total number of measurements required to estimate $\bar{F}_{\mathscr{W}}$ up to error at most $\epsilon > 0$ and failure probability at most $\Delta$ scales as
\begin{equation}
\mathscr{O}\left( \frac{m^{4} s_{\mathscr{U}}^{4} \left[ \Gamma_{\textup{max}} + E_{max}^{\mathscr{U}} \, E_{max}^{\mathscr{E}} \right] \, \ln\left( 2 / \Delta \right) }{\epsilon^{2}} \right), \label{eq:theoremChannels}
\end{equation}
where $s_{\mathscr{U}} \coloneqq \exp(\xi_{max}^{\mathscr{U}})$.
\end{theorem}

This choice of ensemble was made for two reasons. 
First, coherent states are easily accessible in the laboratory, which makes easier the life of a experimentalist trying to certify a real device. 
Second, by choosing a set fully composed of coherent states we arrive at a sample complexity that is \textit{independent of any choice of coherent amplitudes $\bm{\alpha}$}. 
Third, due to the fact that the covariance matrices of $\projector{\bm{0}}$ and $\projector{\bm{\alpha}}$ are equal (both proportional to the $2m \times 2m$ identity matrix $\mathbbm{1}_{2m}$), the sample complexity of our certification protocol \textit{does not depend on the choice of prior distribution}. 
However, it is important to state that different prior distributions will render different values of $\bar{\mathscr{W}}_{\Omega}$, as clearly seen in Eq. \eqref{eq:fidelityChannels}. 
Thus, even though the sample complexity is independent of the ensemble in Eq. \eqref{eq:Omega}, the choices of ensemble and prior distribution are still relevant. 
Furthermore, we note that, since $||\mathbf{x}_{\mathscr{U}}(\ket{\bm{\alpha}})||_{2}^{2} \leq m \, E_{max}^{\mathscr{U}}, \, \forall \, \ket{\bm{\alpha}} \in \mathscr{S}$, Theo. \ref{theorem:Channels} displays the same sample complexity as Theo. \ref{theorem:States}. 
Lastly, if $\mathscr{U}(\projector{\bm{\alpha}})$ is only composed of linear optical elements, then $s_{\mathscr{U}} = 1$. This measurement scheme is detailed in Sec. \ref{SectionIVB}.

We would also like to restate that if the protocol is successful, it is only demonstrated that the experimental gate is close to the ideal gate w.r.t. actions on the input ensemble.
However, we stick to the aforementioned choice of ensemble because using an input ensemble composed of more complex (non-)Gaussian states creates an overhead in complexity that defeats the purpose of requesting the least amount of resources from an experimental perspective (e.g. an overhead of resources and time dedicated to producing and certifying an ensemble of squeezed stated before certifying the action of the experimental gate such ensemble).
The same argument is valid for the ensemble chosen to certify the cubic phase gate in Sec. \ref{SectionIIICII}.

%----------------------------------------------------------------------------------------------
\subsection{Single-mode target channels}
\label{SectionIIIC}
%----------------------------------------------------------------------------------------------

In this subsection we consider two single-mode channel applications: the coherent state amplifier and the cubic phase gate.

\subsubsection{Coherent-state amplifier}
\label{SectionIIICI}

An ideal noiseless amplifier transforms a coherent state $\ket{\alpha}$ into an amplified coherent state $\ket{g \, \alpha}$ as \cite{Zavatta2010, Chiribella2013, Bai2018}
\begin{equation}
\ket{\alpha} \rightarrow \ket{g \, \alpha}, \label{eq:coherentStateAmplifier}
\end{equation}
where $g > 1$ is the gain of the amplifier. However, arbitrary quantum states cannot be perfectly amplified deterministically without the addition of noise \cite{Caves1982}. Transformation \eqref{eq:coherentStateAmplifier} is therefore unphysical if deterministic, but it can be implemented as a non-deterministic (non-unitary) transformation \cite{Ralph2008}. There are several experimental implementations \cite{Ferreyrol2010, Xiang2010} of such probabilistic channel \cite{Xiang2010, Usuga2010, Ferreyrol2010, Zavatta2010, Kocsis2012}.

Here, our goal is to lower-bound the closeness of an experimental implementation $\mathscr{E}$ to the ideal, non-physical transformation \eqref{eq:coherentStateAmplifier}. In order to certify the experimental realization of such channel \cite{Chiribella2013, Bai2018, Namiki2008, Hammerer2005} we can use the fact that, for each experimental run, the state fidelity witness \eqref{eq:stateWitnessHilbert} can be written as
\begin{equation}
W_{g} = \mathbbm{1} - D\left(g \, \alpha\right) \, n \, D^{\dagger}\left(g \, \alpha\right), \label{eq:}
\end{equation}
where $D\left(\alpha\right) \coloneqq \exp(\alpha \, a^{\dagger} - \alpha^{*} \, a)$ is the single-mode displacement operator. We then write the fidelity lower bound $\mathscr{W}_{g}$ as
\begin{eqnarray}
F \geq \mathscr{W}_{g} & \coloneqq & \frac{3}{2} - \tr(\left[\, q - g \, \Re(\alpha)\right]^{2} \, \mathscr{E}(\projector{\alpha})) \nonumber \\ 
& & \, - \tr( \left[\, p - g \, \Im(\alpha)\right]^{2} \mathscr{E}(\projector{\alpha})). \label{eq:fidelityWitnessCoherentAmplifier}
\end{eqnarray}
Writing $\projector{g \, \alpha}$ as the output state of the ideal transformation, we can use \eqref{eq:fidelityWitnessCoherentAmplifier} to lower-bound the average channel fidelity as
\begin{equation}
\bar{F}_{\Omega} \geq \bar{\mathscr{W}}_{\Omega, \, g} \coloneqq \sum_{\ket{\psi} \in \mathscr{S}} \, \p{\alpha} \, \mathscr{W}_{g}, \label{eq:fidelityWitnessAverageCoherentAmplifier}
\end{equation}
where $\bar{\mathscr{W}}_{\Omega, \, g}$ is the average channel-fidelity lower bound for the ideal coherent state amplifier. Analogously to the previous sections, we characterize the number of measurements required to approximate $\bar{\mathscr{W}}_{\Omega, \, g}$ by an estimate $\bar{\mathscr{W}}_{\Omega, \, g}^{*}$ up to $\epsilon$ and $\Delta$. It is assumed here that fourth-order statistical moments are bounded, \textit{i.e.} $r_{\textup{max}} \coloneqq \max_{\ket{\alpha} \in \mathscr{S}} \left\{ \braket{q^{4}}_{\mathscr{E}(\projector{\alpha})} \, , \, \braket{p^{4}}_{\mathscr{E}(\projector{\alpha})} \right\} < \infty$. For the set $\mathscr{S}$ we can define 
\begin{equation}
\mathscr{S}_{\textup{max}} \coloneqq 2 \, \left( 1 + \max_{\ket{\alpha} \in \mathscr{S}} \abs{\Re(\alpha)} + \max_{\ket{\alpha} \in \mathscr{S}} \abs{\Im(\alpha)} \right). \label{eq:boundSet}
\end{equation}
The Theorem below summarizes our results.

\begin{theorem}[Sample complexity of certifying the coherent state amplifier]
\label{theorem:coherentStateAmplifier}
For $g > 1$, let $\mathscr{E}$ be the experimentally implemented coherent-state amplifier channel, with forth-order moments upper-bounded by $r_{\textup{max}}$. In addition, let $\Omega = \left\{  \p{\alpha}, \, \ket{\alpha}\right\}_{\ket{\alpha} \in \mathscr{S}}$ be the probe ensemble used, with bound $\mathscr{S}_{\textup{max}}$ defined in \eqref{eq:boundSet}. Then, the maximum number of measurements required to estimate $\bar{\mathscr{W}}_{\Omega, \, g}$ up to error $\epsilon$ and failure probability $\Delta$ scales as
\begin{equation}
\mathscr{O}\left( \frac{\mathscr{S}_{\textup{max}}^{2} \, r_{\textup{max}} \, \ln(2 / \Delta)}{\epsilon^{2}} \right). \label{eq:theoremCoherentStateAmplifier}
\end{equation}
\end{theorem}

As in the previous cases, \eqref{eq:theoremCoherentStateAmplifier} does not depend on the prior distribution. This measurement scheme is detailed in Sec. \ref{SectionIVCI}.

\subsubsection{Cubic phase gate}
\label{SectionIIICII}

Our framework can also be applied to certify non-Gaussian channels. Here, we show how to certify the non-Gaussian single-mode cubic phase gate, which generates the ideal, unnormalizible \emph{cubic phase state} when applied to the zero-momentum eigenstate $\ket{p = 0}$ \cite{Gottesman2001}. Experimental progress has been made in recent years regarding the preparation of the cubic phase state \cite{Yukawa2013, Miyata2016, Takeda2017}, thus justifying the need for efficient certification protocols for both the cubic phase state \cite{Liu2018} and the cubic phase gate. As stated in Sec. \ref{SectionI}, this is a non-Gaussian element that is sufficient to provide, together with all Gaussian channels, universal quantum computation using continuous-variable degrees of freedom \cite{Gottesman2001, Yukawa2013, Marshall2015, Takeda2017}. For $\gamma \in \mathbb{R}$, the cubic phase gate is defined as
\begin{equation}
U\left(\gamma\right) \coloneqq \exp \left( i \, \gamma \, q^{3} \right) \, . \label{eq:cubicPhaseGate}
\end{equation}
We probe an experimental implementation $\mathscr{E}$ of the cubic phase gate $U\left(\gamma\right)$ with the same ensemble $\Omega$ defined in Sec. \ref{SectionIIICII}, \textit{i.e.} target output states are 
\begin{equation}
\mathscr{U}_{\gamma}\left(\projector{\alpha}\right) = U(\gamma)\projector{\alpha}U^{\dagger}(\gamma). \label{eq:outputTargetCubicPhase} 
\end{equation}
For each experimental realization, the fidelity witness $W_{\gamma}$ is \cite{Liu2018}
\begin{equation}
W_{\gamma} = \mathbbm{1} - U\left(\gamma\right) \, D\left(\alpha\right) \, n \, D^{\dagger}\left(\alpha\right) \, U^{\dagger}\left(\gamma\right). \label{eq:witnessCubicPhaseGate}
\end{equation}
which is demonstrated in App. \ref{appendix:FidelityWitnessCubicPhase}. The state fidelity is then lower-bounded by $\mathscr{W}_{\gamma}$ as
\begin{eqnarray}
F \geq \mathscr{W}_{\gamma} & \coloneqq & \frac{3}{2} - \tr(\left[ 3 \, \gamma \, q^{2} - p + \textup{Im}(\alpha) \right]^{2} \mathscr{E}\left(\projector{\alpha}\right)) \nonumber \\
& & - \, \tr( \left[ q - \textup{Re}(\alpha) \right]^{2} \mathscr{E}\left(\projector{\alpha}\right)) \, , \label{eq:fidelitySingleCubicPhase}
\end{eqnarray}
which is also demonstrated in App. \ref{appendix:FidelityWitnessCubicPhase}. Furthermore, it is demonstrated in the App. \ref{appendix:CubicPhaseGateSampling} how $\mathscr{W}_{\gamma}$ can be expressed in terms of observables that are accessible directly by homodyne detection. Averaging over several experimental runs, we can use Eq. \eqref{eq:fidelitySingleCubicPhase} to lower-bound the average channel fidelity of the process as
\begin{equation}
\bar{F}_{\Omega}\left(\mathscr{U}_{\gamma}\left(\projector{\alpha}\right), \, \mathscr{E}\left(\projector{\alpha}\right) \right) \geq \bar{\mathscr{W}}_{\Omega, \, \gamma} \coloneqq \sum_{\ket{\alpha} \in \mathscr{S}} \p{\alpha} \mathscr{W}_{\gamma}. \label{eq:fidelityCubicPhaseAverage}
\end{equation}
In order to estimate $\bar{\mathscr{W}}_{\Omega, \, \gamma}$ up to $\epsilon$ and $\Delta$, we assume that statistical moments are bounded up to eighth order. Precisely, we assume that $q_{\textup{max}} \coloneqq \max_{\ket{\alpha} \in \mathscr{S}} \tr(q^{8} \, \mathscr{E}(\projector{\alpha})) < \infty$. For the set $\mathscr{S}$ we define the bound
\begin{eqnarray}
\mathscr{S}^{'}_{\textup{max}} & \coloneqq & 1 + \frac{9 \, \gamma^{2}}{4} + \left(1 + 2 \, \sqrt{2}\right) \abs{\gamma} + \max_{\alpha \in \mathscr{S}} \abs{1 + 3 \, \gamma \, \Im(\alpha)} \nonumber \\
& & + \, 2 \, \left( \max_{\alpha \in \mathscr{S}} \abs{\Re(\alpha)} + \max_{\alpha \in \mathscr{S}} \abs{\Im(\alpha)} \right) \label{eq:boundSet2}
\end{eqnarray}
Then, our results for the sample complexity are presented in the Theorem below.

\begin{theorem}[Sample complexity of certifying the cubic phase gate]
\label{theorem:cubicPhaseGate}
For $\gamma \in \mathbb{R}$, let $U(\gamma) \coloneqq \exp(i \, \gamma \, q^{3})$ be the single-mode cubic phase gate, and let $\mathscr{E}$ be the experimentally implemented channel. Let $\Omega = \left\{\p{\alpha}, \, \ket{\alpha}\right\}_{\ket{\alpha} \, \in \, \mathscr{S}}$ be the probe ensemble, with bound $\mathscr{S}^{'}_{\textup{max}}$ defined in \eqref{eq:boundSet2}. Moments of eighth order are assumed to be upper-bounded by $q_{\textup{max}}$. Then, 
\begin{equation}
\label{eq:theoremCubicPhase}
\mathscr{O}\left( \frac{\mathscr{S}^{'\,2}_{\textup{max}} \, q_{\textup{max}} \, \ln(2 / \Delta)}{\epsilon^{2}} \right)
\end{equation}
is the maximum number of measurements required to estimate $\mathscr{W}_{\gamma}$ up to error $\epsilon$ and failure probability $\Delta$.
\end{theorem}

As it was true for the estimation of average channel-fidelity witnesses of Gaussian unitary channels and the coherent state amplifier, choosing coherent states as input yields a sample complexity that does not depend on the choice of prior distribution. Section \ref{SectionIVCII} presents the respective measurement scheme.
 
%%%%%%%%%%%%%%%%%%%%%%%%%%%%%%%%%%%%%%%%%%%%%%%%%%%%%%%%%%%%%%%%%%%%%%%%%%%%%%%%%%%%%%%%%%%%%%%
\section{Measurement Schemes}
\label{SectionIV}
%%%%%%%%%%%%%%%%%%%%%%%%%%%%%%%%%%%%%%%%%%%%%%%%%%%%%%%%%%%%%%%%%%%%%%%%%%%%%%%%%%%%%%%%%%%%%%%

In the following subsections we show how to use importance sampling techniques to estimate the fidelity lower bounds presented in Sec. \ref{SectionIII}. For clearance, we present the measurement scheme for pure Gaussian target states before the measurement scheme for Gaussian unitary target channels. Then, we present the measurement scheme for the single-mode applications.

%----------------------------------------------------------------------------------------------
\subsection{Pure Gaussian target states}
\label{SectionIVA}
%----------------------------------------------------------------------------------------------

We start from \eqref{eq:fidelityStates}. Note that the third overlap term on its right-hand side (\textit{r.h.s.}) is a known quantity, since $\mathbf{V}_{t}$ and $\mathbf{x}_{t}$ are known. Thus, we must estimate the remaining overlaps, $\Tr(\mathbf{V}_{t}^{-1} \, \mathbf{x}_{p} \, \mathbf{x}_{t}^{T})$ and $\Tr(\mathbf{V}_{t}^{-1} \, \bm{\Gamma}_{p})$. Each overlap is written as the expectation value of an \textit{estimator} which is dependent on quadrature measurements and associated with a joint probability distribution. By sampling relevant quadratures from the joint probability distribution and measuring them via homodyne detection, one can directly obtain the expectation value of the estimator, and consequently obtain an estimation for each overlap without full knowledge of $\bm{\Gamma}_{p}$ and $\mathbf{x}_{p}$. Next subsections we define the two estimators suitable to estimate each overlap as well as their associated probability distribution $\textup{p}$.

\subsubsection{First-moment estimator}
\label{SectionIVAI}
Let us first focus on the overlap involving first-moment vectors. We can define a probability distribution $\p{k,\,l}$ such that
\begin{equation}
\p{k,\,l} = \frac{\left[\mathbf{V}_{t}^{-1}\right]_{kl}^{2}}{ \left|\left|\mathbf{V}_{t}^{-1}\right|\right|_{F}^{2}}. \label{eq:probabilityKL}
\end{equation}
where $||\mathbf{V}_{t}^{-1}||_{F}^{2} = \Tr(\mathbf{V}_{t}^{-2}) = \sum_{k,l} \left[\mathbf{V}_{t}^{-1}\right]_{kl}^{2}$ is the squared \textit{Frobenius norm} of $\mathbf{V}_{t}^{-1}$. Thus, we use the definition of trace operation to write 
\begin{equation}
\Tr(\mathbf{V}_{t}^{-1} \, \mathbf{x}_{p} \, \mathbf{x}_{t}^{T}) = \sum_{k,l} \, \p{k, \, l} \, \frac{\left[\mathbf{x}_{p}\right]_{k} \, \left[\mathbf{x}_{t}^{T}\right]_{l}}{\left[\mathbf{V}_{t}^{-1}\right]_{kl}} \, \left|\left|\mathbf{V}_{t}^{-1}\right|\right|_{F}^{2}. \label{eq:firstMomentsFirstStep}
\end{equation}
where we have used the fact that $\left[\mathbf{V}_{t}^{-1}\right]_{kl} \neq 0$ for all $\{k, l\}$ relevant to the summation. Since each element of $\mathbf{V}_{t}$ is associated with a bilinear combination of field quadratures, each pair $(k,l)$ corresponds to observables that can be measured with homodyne detection. Each target state defines a fixed prior distribution $\p{k,\,l}$ and it defines the observables that are most relevant to the estimation of $\Tr(\mathbf{V}_{t}^{-1} \, \mathbf{x}_{p} \, \mathbf{x}_{t}^{T})$. Furthermore, from the definition of first-moment vectors presented in \eqref{eq:firstMoments}, $[\mathbf{x}_{p}]_{k}$ can be expressed as
\begin{equation}
\left[\mathbf{x}_{p}\right]_{k} = \left[\braket{ \mathbf{r} }_{\varrho_{p}}\right]_{k} = \int \, dr' \, \p{r' \, | \, k} \, r'_{k}, \label{eq:firstMomentsIntegral}
\end{equation}
where $\p{r' \, | \, k} = \tr(\Pi_{k, r'} \, \varrho_{p})$ is the probability distribution of measuring eigenvalue $r' \in \mathbb{R}$ given the $k$-th observable ($q_{k}$ if $k$ is odd, $p_{k}$ if $k$ is even), and $\Pi_{k, r'}$ is the projector onto the eigenstate with eigenvalue $r'$ of the $k$-th quadrature measurement. Now, substituting \eqref{eq:probabilityKL} and \eqref{eq:firstMomentsIntegral} into \eqref{eq:firstMomentsFirstStep}, we have
\begin{equation}
\Tr(\mathbf{V}_{t}^{-1} \, \mathbf{x}_{p} \, \mathbf{x}_{t}^{T}) = \sum_{k,l} \int dr' \, \p{k, \, l, \, r'} \, \chi_{k, \, l, \, r'} = \mathbb{E}\left( \chi \right), \label{eq:firstMomentsAverageEstimator}
\end{equation}
where we defined the \textit{estimator} $\chi$ with possible values
\begin{equation}
\chi_{k, \, l, \, r'} \coloneqq r'_{k} \, \frac{\left[\mathbf{x}_{t}^{T}\right]_{l}}{\left[\mathbf{V}_{t}^{-1}\right]_{kl}} \, \left|\left|\mathbf{V}_{t}^{-1}\right|\right|_{F}^{2} \, . \label{eq:firstMomentsEstimator}
\end{equation}
Notation $\mathbb{E}(\cdot)$ denotes the expectation value of a random variable, and $\p{k, \, l, \, r'} = \p{r' \, | \, k} \, \p{k, \, l}$ is the joint probability of measuring observables corresponding to pair $\{k,\,l\}$ and obtaining $r'$ as result.

Equation \eqref{eq:firstMomentsAverageEstimator} tells us that it is possible to understand the overlap $\Tr(\mathbf{V}_{t}^{-1} \, \mathbf{x}_{p} \, \mathbf{x}_{t}^{T})$ as the expectation value of a single unbounded random variable $\chi$. Even though each possible value $\chi_{k, \, l, \, r'}$ is accessible by multiplying measurement result $r'_{k}$ by its associated constant $\left[\mathbf{x}_{t}^{T}\right]_{l} \left|\left|\mathbf{V}_{t}^{-1}\right|\right|_{F}^{2} / \left[\mathbf{V}_{t}^{-1}\right]_{kl} \,$, an infinite number of measurements would be necessary to estimate $\mathbb{E}(\chi)\,$, as $r' \in \mathbb{R}\,, \forall \, k \in [2m]$. In App. \ref{SectionVAI} we show the sample complexity of estimating $\mathbb{E}\left( \chi \right)$, which is part of the result in Theo. \ref{theorem:States}.

\subsubsection{Second-moment estimator}
\label{SectionIVAII}
Again, it follows from the definition of the trace operation that
\begin{equation}
\Tr(\mathbf{V}_{t}^{-1} \, \bm{\Gamma}_{p}) = \sum_{k,\,l} \, \p{k, \, l} \, \frac{\left[\bm{\Gamma}_{p}\right]_{kl}}{\left[ \mathbf{V}_{t}^{-1} \right]_{kl}} \, \left|\left|\mathbf{V}_{t}^{-1}\right|\right|_{F}^{2} \, , \label{eq:secondMomentsFirstStep}
\end{equation}
where $\p{k, \, l}$ was defined in \eqref{eq:probabilityKL}. Analogously to \eqref{eq:firstMomentsIntegral}, we can use \eqref{eq:secondMoments} to write each matrix element $\left(\bm{\Gamma}_{p}\right)_{kl}$ as
\begin{equation}
\left[\bm{\Gamma}_{p}\right]_{kl} = \int \, d\Gamma' \, \p{\Gamma' \, | \, k, \, l} \, \Gamma_{kl}' \, , \label{eq:secondMomentsIntegral}
\end{equation}
where $\p{\Gamma' \, | \, k, \, l} = \tr(\Pi_{k, \, l, \, \Gamma'} \, \varrho_{p})$ is the probability distribution of measuring observales given by the pair $(k,l)$ and obtaining eigenvalue $\Gamma_{kl}' \in \mathbb{R}$. It is possible to write the r.h.s. of \eqref{eq:secondMomentsFirstStep} as the expectation value of the estimator $X$, \textit{i.e.}
\begin{equation}
\Tr(\mathbf{V}_{t}^{-1} \, \bm{\Gamma}_{p}) = \sum_{k, \, l} \, \int d\Gamma' \, \p{k, \, l, \, \Gamma'} \, X_{k,l,\Gamma'} = \mathbb{E}\left( X \right), \label{eq:secondMomentsEstimator}
\end{equation}
where $X$ can assume the possible values
\begin{equation}
X_{k, \, l, \, \Gamma'} \coloneqq \frac{\Gamma_{kl}'}{\left[ \mathbf{V}_{t}^{-1} \right]_{kl}} \, \left|\left|\mathbf{V}_{t}^{-1}\right|\right|_{F}^{2} \, , \label{eq:secondMomentsStatesEstimator}
\end{equation}
and $\p{k, \, l, \, \Gamma'} = \p{\Gamma' \, | \, k, \, l} \, \p{k, \, l}$. Since $\bm{\Gamma}_{p}$ is a $2m \times 2m$ matrix composed of $4 m^{2}$ expectation values with $2 m$ of them involving non-commuting observables, in App. \ref{appendix:secondMomentsMeasurementScheme} we go through a discussion about how to measure $\Gamma_{kl}'$. In App. \ref{SectionVAII} we show the sample complexity of estimating $\mathbb{E}\left( X \right)$, which complements the results in App. \ref{SectionVAI} in order to prove Theo. \ref{theorem:States}.

\textit{\textbf{Protocol 1}}: To illustrate how we envision the measurement protocol working in practice, we give present step-by-step heuristics as follows:

\begin{enumerate}

\item One defines a target pure state $\varrho_{t}$. One wants to certify how close the experimental implementation $\varrho_{p}$ is from $\varrho_{t}$;

\item Defining an error $\epsilon$, a probability of success $1 - \Delta$, and upper-bounds for single-mode energy $E_{\text{max}}^{(p)}$ and fourth-order moments $\Gamma_{\text{max}}$, the experimentalist knows that it will need to accumulate a number of measurement results that at most equates Eq. \eqref{eq:theoremStates};

\item From $\varrho_{t}$ one is able to sample observables according to the prior distribution $\text{p}(k, \, l)$ in order to measure via homodyne detection;

\item From the experimental statistics accumulated, the experimentalist is able to obtain eigenvalues $\Gamma_{kl}'$ according to the conditional distribution $\text{p}(\Gamma' \, | \, k, \, l)$ for second moments as well as eigenvalues $r_{k}'$ according to $\text{p}(r' \, | \, k)$ for first moments;

\item At last, the experimentalist classicaly process the $N$ measurement results according to median-of-means protocol and evaluate the estimators $X$ and $\chi$ defined, respectively, in Eqs. \eqref{eq:secondMomentsEstimator} and \eqref{eq:firstMomentsAverageEstimator} (see App. \ref{appendix:SectionV}).
\end{enumerate}

Since the step-by-steps from the above protocol changes very little for all the other case studies mentioned in this pap1er, from now own we refer to this Section when it comes to the measurement scheme's pseudo-algorithm.

%----------------------------------------------------------------------------------------------
\subsection{Gaussian unitary target channels}
\label{SectionIVB}
%----------------------------------------------------------------------------------------------
There are two terms in Eq. \eqref{eq:fidelityChannels} that are dependent of the experimental channel. For the overlap involving first moments, it is possible to define the estimator $\chi^{(c)}\,$,  similarly to Eq. \eqref{eq:firstMomentsEstimator}, with possible values
\begin{equation}
\chi_{k, l, r', \psi}^{(c)} \coloneqq r_{k}'(\ket{\psi}) \, \frac{ \left[\mathbf{x}_{\mathscr{U}}^{T}(\ket{\psi})\right]_{l}}{\left[\mathbf{V}_{\mathscr{U}}^{-1}(\ket{\psi})\right]_{kl}} \left|\left|\mathbf{V}_{\mathscr{U}}^{-1}(\ket{\psi})\right|\right|_{F}^{2} \, , \label{eq:firstMomentsEstimatorChannel}
\end{equation}
where there are implicit dependencies on the set of input states, $\mathscr{S} = \{\ket{\psi}\}$. From now on, the superscript $(c)$ refers to channels, and we do not explicitly write the dependencies on $\mathscr{S}$ for simplicity. Nevertheless, the second integral on the r.h.s. of Eq. \eqref{eq:fidelityChannels} can be rewritten as
\begin{equation}
\sum_{\ket{\psi} \in \mathscr{S}} \, \p{\psi} \Tr\left[ \mathbf{V}_{\mathscr{U}}^{-1} \mathbf{x}_{\mathscr{E}} \mathbf{x}_{\mathscr{U}}^{T}\right] = \mathbb{E}\left( \chi^{(c)} \right) \, , \label{eq:firstMomentsAverageEstimatorChannels}
\end{equation}
where
\begin{equation}
\mathbb{E}\left(\chi^{(c)}\right) = \sum_{k,\,l,\,\mathscr{S}} \, \int dr' \, \p{k,\,l,\,r',\,\psi} \, \chi_{k,l,r',\psi}^{(c)} \, , \label{eq:firstMomentsAverageEstimatorChannels2}
\end{equation}
and $\p{k, \, l, \, r', \, \psi} = \p{\psi} \, \p{k, \, l, \, r' \, | \, \psi} \, $. We see that $\mathbb{E}\left(\chi^{(c)}\right)$ depends on the prior distribution $\p{\psi}$. As for the estimation of second moments, the estimator $X^{(c)}$ is defined as in Eq. \eqref{eq:secondMomentsStatesEstimator}, but now with implicit dependencies on input states $\ket{\psi}$, with possible values
\begin{equation}
X_{k, \, l, \, \Gamma', \, \psi} \coloneqq \frac{\Gamma_{kl}'(\ket{\psi})}{\left[ \mathbf{V}_{t}^{-1}(\ket{\psi}) \right]_{kl}} \, \left|\left|\mathbf{V}_{t}^{-1}(\ket{\psi})\right|\right|_{F}^{2} \, . \label{eq:secondMomentsChannelsEstimator}
\end{equation}
Thus, the third mean overlap on the r.h.s. of \eqref{eq:fidelityChannels} can be rewritten as
\begin{equation}
\sum_{\mathscr{S}} \, \p{\psi} \, \Tr[\mathbf{V}_{\mathscr{U}}^{-1} \, \bm{\Gamma}_{\mathscr{E}}] = \mathbb{E}\left(X^{(c)}\right) \, , \label{eq:secondMomentsAverageEstimatorChannels}
\end{equation}
where
\begin{equation}
\mathbb{E}\left(X^{(c)}\right) = \sum_{k,\,l,\,\mathscr{S}} \, \int d\Gamma' \, \p{k, \, l, \, \Gamma', \, \psi} \, X_{k, \, l, \, \Gamma', \, \psi}^{(c)} \, , \label{eq:secondMomentsAverageEstimatorChannels2}
\end{equation}
and $\p{k, \, l, \, \Gamma', \, \psi} = \p{\psi} \, \p{k, \, l, \, \Gamma' \, | \, \psi}\,$. Here, we see that $\mathbb{E}\left(X^{(c)}\right)$ also depends on $\p{\psi}\,$, as expected. In App. \ref{SectionVB} we show how the estimation of $\mathbb{E}\left( \chi^{(c)} \right)$ and $\mathbb{E}\left( X^{(c)} \right)$ leads to the sample complexity in Theo. \ref{theorem:Channels}.

%----------------------------------------------------------------------------------------------
\subsection{Single-mode target channels}
\label{SectionIVC}
%----------------------------------------------------------------------------------------------

Here we state the measurement schemes for the two single-mode applications considered in Sec. \ref{SectionIIIC}.

\subsubsection{Coherent-state amplification channel}
\label{SectionIVCI}

By combining \eqref{eq:fidelityWitnessCoherentAmplifier} and \eqref{eq:fidelityWitnessAverageCoherentAmplifier}, the average channel-fidelity lower-bound $\bar{\mathscr{W}}_{\Omega, \, g}$ can be written as 
\begin{align}
\label{eq:fidelityCoherentStateAmplifierSampling}
\nonumber
\bar{\mathscr{W}}_{\Omega, \, g} & = \frac{3}{2} - g^{2} \sum_{\ket{\alpha} \in \mathscr{S}} \, \p{\alpha} \abs{\alpha}^{2} \\
& \qquad \, - \mathop{\sum_{\ket{\alpha} \in \mathscr{S}}}_{k} \, \p{\alpha} \tau_{k} \, \tr(\nu_{k} \, \mathscr{E}\left(\projector{\alpha}\right)), 
\end{align}
where sets of observables $\{\nu_{k}\}$ and coefficients $\{\tau_{k}\}$ are such that
\begin{equation}\label{eq:observablesCoherentStateAmplifier}
\begin{array}{r@{}l}
\left\{\nu_{1},\nu_{2},\nu_{3},\nu_{4}\right\} &{}\coloneqq \{ q^{2}, \, p^{2}, \, q, \, p \} \, ; \\
\left\{\tau_{1},\tau_{2},\tau_{3},\tau_{4}\right\} &{}\coloneqq \{ 1, \, 1, \, - 2 \, g \, \Re(\alpha), \, - 2 \, g \, \Im(\alpha) \} \, .
\end{array}
\end{equation}
Since the first sum in \eqref{eq:fidelityCoherentStateAmplifierSampling} is a completely defined by the ensemble $\Omega$, the certification of the coherent state amplifier relies on the estimation of the four expectation values from the remaining sum. We show below how to write \eqref{eq:fidelityCoherentStateAmplifierSampling} as the expectation value of a single unbounded random variable. First, we can rewrite each $\tau_{k}$ as $\textup{sign}(\tau_{k}) \, \abs{\tau_{k}}$, where $\textup{sign}(\tau_{k}) = 1$ if $\tau_{k} \geq 0$, and $-1$ otherwise. We can also define the probability distribution $\p{k \, | \, \alpha}$ as
\begin{equation}
\p{k \, | \, \alpha} \coloneqq \frac{\abs{\tau_{k}(\alpha)}}{\sum_{l} \abs{\tau_{l}(\alpha)}} \, . \label{eq:probabilityKAmplifier}
\end{equation}
Each quadrature operator $\nu_{k}$ can be expressed in its respective diagonal basis as
\begin{equation}
\nu_{k} = \int d\nu' \, \Pi_{k, \nu'} \, \nu'_{k} \, , \label{eq:coherentStateAmplifierMomentsIntegral}
\end{equation}
where $\Pi_{k, \nu'}$ is the projection of the $k$-th quadrature operator onto the eigenstate with eigenvalue $\nu'$. We can also define
\begin{equation}
\p{ \nu' \, | \, k, \, \alpha} \coloneqq \tr( \Pi_{k, \nu'} \, \mathscr{E}\left(\projector{\alpha}\right) ) \label{eq:probabilityConditionalKFCoherentStateAmplifier}
\end{equation}
as the conditional probability distribution of, given input state $\ket{\alpha}$, choosing to measure the $k$-th observable and obtaining eigenvalue $\nu'$ as result. Thus, $\p{k, \, \nu', \, \alpha} = \p{\alpha} \, \p{k \, | \, \alpha} \, \p{\nu' \, | \, k, \, \alpha}$. As in the case of Gaussian unitary target channels, an estimator $\zeta$, with possible values
\begin{equation}
\zeta_{k, \, \nu', \, \alpha} \coloneqq \textup{sign}\left(\tau_{k}(\alpha)\right) \, \nu' \, \sum_{l} \abs{\tau_{l}(\alpha)} \, , \label{eq:coherentStateAmplifierEstimator}
\end{equation}
can be defined. We rewrite the second sum on the r.h.s. of \eqref{eq:fidelityCoherentStateAmplifierSampling} as
\begin{equation}
\sum_{k,\,\mathscr{S}} \, \p{\alpha} \, \tau_{k} \, \tr\left(\nu_{k} \, \mathscr{E}\left(\projector{\alpha}\right) \right) = \mathbb{E}\left( \zeta \right) \, , \label{eq:coherentStateAmplifierMean}
\end{equation}
where
\begin{equation}
\mathbb{E}(\zeta) \coloneqq \sum_{k,\,\mathscr{S}} \, \int \, d\nu' \, \p{k, \, \nu', \, \alpha} \, \zeta_{k, \, \nu', \, \alpha} \, . \label{eq:ExpectationValueZeta}
\end{equation}
In App. \ref{SectionVC} we show that the estimation of $\mathbb{E}(\zeta)$ yields the sample complexity in Theo. \ref{theorem:coherentStateAmplifier}.

\subsubsection{Cubic phase gate}
\label{SectionIVCII}

Here, we follow the same mathematical steps as in Sec. \ref{SectionIVCI}. From \eqref{eq:fidelitySingleCubicPhase} and \eqref{eq:fidelityCubicPhaseAverage}, the average channel-fidelity lower-bound $\bar{\mathscr{W}}_{\Omega, \, \gamma}$ can be rewritten as
\begin{align}
\label{eq:fidelityCubicPhaseGateSampling}
\nonumber
\bar{\mathscr{W}}_{\Omega, \, \gamma} & = \frac{3}{2} - \sum_{\ket{\alpha} \in \mathscr{S}} \p{\alpha} \abs{\alpha}^{2} \\
& \qquad \, - \mathop{\sum_{\ket{\alpha} \in \mathscr{S}}}_{k} \p{\alpha} \, \kappa_{k} \, \tr(\mu_{k} \, \mathscr{E}\left(\projector{\alpha}\right)),  
\end{align}
where the sets of coefficients $\{\kappa_{k}\}$ and observables $\{\mu_{k}\}$ are such that
\begin{equation}
\begin{array}{r@{}l}
\left\{\mu_{1}, \cdots, \mu_{8}\right\} &{}\coloneqq \{ q^{4}, \, \left((q + p) / \sqrt{2}\right)^{3}, \, \left((q - p) / \sqrt{2}\right)^{3}, \label{eq:observablesCubicPhase} \\
&{} \quad \quad \quad \, p^{3}, \, q^{2}, \, p^{2}, \, q, \, p \} ; \\
\\
\left\{\kappa_{1}, \cdots, \kappa_{8}\right\} &{}\coloneqq \{ 9 \, \gamma^{2} / 4, \, \sqrt{2 } \, \gamma, \, - \sqrt{2} \, \gamma , \, - \gamma, \\ 
&{} \quad \quad \quad \, - \left(1 + 3 \, \gamma \, \Im(\alpha)\right), \, - 1 , \\
&{} \quad \quad \quad \, - 2 \, \Re(\alpha), \, - 2 \, \Im(\alpha) \} \, .
\end{array}
\end{equation}
Now, the conditional probability distribution $\p{k \, | \, \alpha}$ is a function of the coefficients $\kappa_{k}$:
\begin{equation}
\p{k \, | \, \alpha} \coloneqq \frac{\abs{\kappa_{k}(\alpha)}}{\sum_{l} \abs{\kappa_{l}(\alpha)}} \, . \label{eq:probabilityKCubicPhase}
\end{equation}
Analogously to \eqref{eq:coherentStateAmplifierMomentsIntegral}, each quadrature operator $\mu_{k}$ can expressed in its respective diagonal basis as
\begin{equation}
\mu_{k} = \int \, d\mu' \, \Pi_{k, \mu'} \, \mu'_{k} \, , \label{eq:cubicPhaseGateMomentsIntegral}
\end{equation}
and the conditional probability distribution $\p{ \mu' \, | \, k, \, \alpha}$ is such that
\begin{equation}
\p{ \mu' \, | \, k, \, \alpha} \coloneqq \tr( \Pi_{k, \mu'} \, \mathscr{E}\left(\projector{\alpha}\right) ) \, . \label{eq:probabilityConditionalKFCubicPhase}
\end{equation}
The joint probability of probing the experimental channel with coherent state $\ket{\alpha}$, choosing to measure observable $k$, and having $\mu'$ as measurement result is therefore $\p{k, \, \alpha, \, \mu'} = \p{\alpha} \, \p{k \, | \, \alpha} \, \p{\mu' \, | \, k, \, \alpha}$. Defining the estimator $Z$ with possible values
\begin{equation}
Z_{k, \, \mu', \, \alpha} \coloneqq \textup{sign}\left(\kappa_{k}(\alpha)\right) \, \mu' \, \sum_{l} \abs{\kappa_{l}(\alpha)}, \label{eq:cubicPhaseGateEstimator}
\end{equation}
we can finally rewrite the second sum on the r.h.s. of \eqref{eq:fidelityCubicPhaseGateSampling} as
\begin{equation}
\mathop{\sum_{\ket{\alpha} \in \mathscr{S}}} \, \p{\alpha} \, \sum_{k} \, \kappa_{k} \, \tr\left(\mu_{k} \, \mathscr{E}\left(\projector{\alpha}\right) \right) = \mathbb{E}\left( Z \right) \, , \label{eq:cubicPhaseGateMean}
\end{equation}
where
\begin{equation}
\mathbb{E}(Z) \coloneqq \mathop{\sum_{k, \, \ket{\alpha}\in\mathscr{S}}}\, \int \, d\mu' \, \p{k, \, \alpha, \, \mu'} \, Z_{k, \, \mu', \, \alpha} \, . \label{eq:ExpectationValueZ}
\end{equation}
In App. \ref{SectionVC} we show that the estimation of $\mathbb{E}(Z)$ yields the sample complexity in Theo. \ref{theorem:cubicPhaseGate}.

%%%%%%%%%%%%%%%%%%%%%%%%%%%%%%%%%%%%%%%%%%%%%%%%%%%%%%%%%%%%%%%%%%%%%%%%%%%%%%%%%%%%%%%%%%%%%%%
\section{Conclusions and Outlook}
\label{SectionVI}
%%%%%%%%%%%%%%%%%%%%%%%%%%%%%%%%%%%%%%%%%%%%%%%%%%%%%%%%%%%%%%%%%%%%%%%%%%%%%%%%%%%%%%%%%%%%%%%

We derived efficiently-measurable witnesses for the average channel fidelity between an arbitrary, unknown experimental gate and an ideal, known gate, for three important classes of target gates: multi-mode Gaussian unitary channels, single-mode coherent state amplifiers, and the single-mode (non-Gaussian) cubic phase gate. Our witnesses are  experimentally-friendly in that they can be measured by probing the channel with simple Gaussian states (coherent states) as inputs and making Gaussian measurements (homodyne detection)  at the  output, even for the non-Gaussian case considered. Moreover, in all three cases, the estimation of their expectation value is efficient in all relevant parameters: its \textit{sample complexity} (i.e. the total number of measurements required) scales polynomially in the inverse estimation error $1/\epsilon$ and logarithmically in the inverse failure probability $1/\Delta$ of the estimation, as well as polynomially in the number of modes $m$ for the Gaussian-target case. Such scaling in $\Delta$ represents an  exponential improvement with respect to previous estimation methods \cite{Aolita2015}.

For the case  Gaussian unitary target channels, our channel-fidelity witness exploits state-fidelity witnesses for Gaussian target states \cite{Aolita2015}. Interestingly, to measure the latter, we develop an enhanced method, based on importance sampling \cite{Gluza2018}, which (apart from the already mentioned improvement on $\Delta$) is polynomially more efficient in $m$ than previous methods \cite{Aolita2015}.  Furtheremore, the resulting sample complexity for the certification of $m$-mode Gaussian target states displays the same scaling as the estimation of the average channel-fidelity lower bound for arbitrary Gaussian unitary channels probed by $m$-mode coherent states. This is a byproduct technical contribution interesting in its own right for certifying state preparations (instead of channels).

Our findings are relevant for the certification of experimental many-body quantum technologies in the continuous-variable domain. Particularly promising prospects may for instance be the certification of the forthcoming first non-Gaussian resources, such as single mode non-Gaussian states and channels, with important implications \cite{Liu2018} for universal quantum computing.

%%%%%%%%%%%%%%%%%%%%%%%%%%%%%%%%%%%%%%%%%%%%%%%%%%%%%%%%%%%%%%%%%%%%%%%%%%%%%%%%%%%%%%%%%%%%%%%
\section{Acknowledgements}
\label{SectionVII}
%%%%%%%%%%%%%%%%%%%%%%%%%%%%%%%%%%%%%%%%%%%%%%%%%%%%%%%%%%%%%%%%%%%%%%%%%%%%%%%%%%%%%%%%%%%%%%%

We acknowledge financial support from the Brazilian agencies CNPq (PQ grant No. 311416/2015-2 and INCT-IQ), FAPERJ (JCNE- 26/202.701/2018), CAPES (PROCAD2013), and the Serrapilheira Institute (grant number Serra-1709-17173).

%%%%%%%%%%%%%%%%%%%%%%%%%%%%%%%%%%%%%%%%%%%%%%%%%%%%%%%%%%%%%%%%%%%%%%%%%%%%%%%%%%%%%%%%%%%%%%%

%\bibliographystyle{apsrev4-1}
\bibliography{Master_Bibtex}

%%%%%%%%%%%%%%%%%%%%%%%%%%%%%%%%%%%%%%%%%%%%%%%%%%%%%%%%%%%%%%%%%%%%%%%%%%%%%%%%%%%%%%%%%%%%%%%

\appendix

%%%%%%%%%%%%%%%%%%%%%%%%%%%%%%%%%%%%%%%%%%%%%%%%%%%%%%%%%%%%%%%%%%%%%%%%%%%%%%%%%%%%%%%%%%%%%%%
\section{Sample Complexities and Proofs of Theorems}
\label{appendix:SectionV}
%%%%%%%%%%%%%%%%%%%%%%%%%%%%%%%%%%%%%%%%%%%%%%%%%%%%%%%%%%%%%%%%%%%%%%%%%%%%%%%%%%%%%%%%%%%%%%%

In this appendix we show how to use the median-of-means estimation protocol in order to estimate all the expectation values defined in Sec. \ref{SectionIV} and prove the sample complexities presented in Sec. \ref{SectionIII}. 

First, we define the general framework of the median-of-means protocol \cite{Nemirovsky1983, Jerrum1986}. 
Let $Y$ be a random variable with variance $\sigma^{2} = \mathbb{E}(Y^{2}) - \mathbb{E}^{2}(Y)$, and let $\epsilon > 0$ be the estimation error. 
We divide the $N$ experimental data points in $B$ batches of size $\floor{N / B}$ such that each batch $\omega$ has $N_{(\omega)} = 34 \, \sigma^{2} / \epsilon^{2}$ independent copies of Y.
Note that $N = B \, N_{(\omega)}$. 
Then, we calculate the empirical mean $\mathbb{E}^{(\omega)}(Y)$ for every batch as
\begin{equation}
\mathbb{E}^{(\omega)}\left(Y\right) = \frac{1}{\floor{N / B}} \, \sum_{j = (\omega - 1)\floor{N / B} + 1}^{\omega \, \floor{N / B}} \, Y_{\bm{y}_{j}} \, , \label{eq:expectationOmega}
\end{equation}
where $\omega \in \{1, 2, \cdots, B\}$, and $\bm{y}_{j}$ represents the parameters of the $j$-th experimental realization.
The median-of-means estimator $\mathbb{E}_{\textup{MM}}(Y)$ is then computed by calculating the median empirical mean among all batches, i.e.
\begin{equation}
\mathbb{E}_{\textup{MM}}\left(Y\right) = \textup{median}\left\{ \mathbb{E}^{(1)}\left(Y\right), \cdots, \mathbb{E}^{(B)}\left(Y\right) \right\}. \label{eq:expectationMM}
\end{equation}
By using the Chernoff inequality \cite{Chernoff1952}, one can show (see, e.g., Ref. \cite{Huang2020}) that the probability the median-of-means estimator diverges from $\mathbb{E}(Y)$ by more than $\epsilon$ follows
\begin{equation}
\mathbb{P}( \, | \, \mathbb{E}\left(Y\right) - \mathbb{E}_{\textup{MM}} \left(Y\right) \, | > \epsilon) \leq 2 \, e^{-B / 2} \eqqcolon \Delta \, , \label{eq:probabilityMM}
\end{equation}
where $\Delta$ is the maximum failure probability.
Hence, using the fact that $\sigma^{2} \leq \mathbb{E}(Y^{2})$, we are able to write $N$ as
\begin{equation}
N = \mathscr{O}\left( \frac{\mathbb{E}(Y^{2}) \, \ln(2 / \Delta)}{\epsilon^{2}} \right) \, . \label{eq:scalingN}
\end{equation}

In the following subsections we apply the results above to each random variable defined in Sec. \ref{SectionIV}, to prove the theorems in Sec. \ref{SectionIII}.

%----------------------------------------------------------------------------------------------
\subsection{Pure Gaussian target states}
\label{SectionVA}
%----------------------------------------------------------------------------------------------

\subsubsection{First-moment estimator}
\label{SectionVAI}

In Sec. \ref{SectionIVAI}, $\Tr(\mathbf{V}_{t}^{-1} \, \mathbf{x}_{p} \, \mathbf{x}_{t})$ was written as the expectation value $\mathbb{E}\left(\chi\right)$, which is measurement-dependent. However, a certifier is constrained by a finite number of state preparations, \textit{i.e.} by a finite number of measurements. What is actually accessible to the certifier is $\mathbb{E}_{\textup{MM}}(\chi)$, which is an empirical estimation of $\mathbb{E}(\chi)$ and is defined as
\begin{equation}
\mathbb{E}_{\textup{MM}}\left(\chi\right) \coloneqq \textup{median}\left\{\mathbb{E}^{(1)}(\chi), \cdots, \mathbb{E}^{(B)}(\chi)\right\} \, , \label{eq:firstMomentsEmpiricalAverage}
\end{equation}
with $B$ being the number of batches, and
\begin{equation}
\mathbb{E}^{(\omega)}\left( \chi \right) \coloneqq \frac{1}{\floor{\mathscr{N}_{1} / B}} \, \sum_{j = (\omega - 1)\floor{\mathscr{N}_{1} / B} + 1}^{\omega \, \floor{\mathscr{N}_{1} / B}} \, \chi_{\bm{\sigma}_{j}} \, ,
\end{equation}
where $\bm{\sigma}_{j} = \{ k_{j}, \, l_{j}, \, r'_{j} \}$ is the $j$-th experimental realization, and $\mathscr{N}_{1}$ is the total number of measurements required to estimate $\mathbb{E}\left(\chi\right)$ with statistical confidence. Now, from Eq. \eqref{eq:probabilityMM} gives us an upper bound $\Delta$ for the probability $\mathbb{P}$ that $\mathbb{E}(\chi)$ and $\mathbb{E}_{\textup{MM}}(\chi)$ differ by more than an error $\epsilon > 0$. As mentioned previously, the upper bound $\Delta$ is called the maximum failure probability. Equation \eqref{eq:scalingN} enables us to write $\mathscr{N}_{1}$ in terms of $\epsilon$ and $\Delta$ as
\begin{equation}
\mathscr{N}_{1} = \mathscr{O}\left(\frac{ \mathbb{E}(\chi^{2}) \, \ln\left(2 / \Delta \right)}{\epsilon^{2}}\right), \label{eq:firstMomentsNumberOfCopies}
\end{equation}
Equation \eqref{eq:firstMomentsNumberOfCopies} already displays the scaling in $\epsilon$ and $\Delta$ presented in Theo. \ref{theorem:States}. In App. \ref{appendix:StatesFirst} we demonstrate how to upper-bound $\mathbb{E}(\chi^{2})$ by
\begin{equation}
\mathbb{E}\left(\chi^{2}\right) \leq 2^{6} \, m^{3} \, s_{t}^{4} \, E_{max}^{(p)} \, \left|\left| \mathbf{x}_{t} \right|\right|_{2}^{2} \, , \label{eq:firstMomentsUpperBoundExpChi2}
\end{equation}
where $s_{t} \coloneqq \exp(\xi_{max}^{(t)})$ and $E_{max}^{(p)}$ was defined in Sec. \ref{SectionIIIA}. Thus, substituting Eq. \eqref{eq:firstMomentsUpperBoundExpChi2} into Eq. \eqref{eq:firstMomentsNumberOfCopies}, we see that
\begin{equation}
\mathscr{N}_{1} = \mathscr{O} \left( \frac{m^{3} \, s_{t}^{4} \, E_{max}^{(p)} \, \left|\left| \mathbf{x}_{t} \right|\right|_{2}^{2} \, \ln(2 / \Delta)}{\epsilon^{2}} \right) \, , \label{eq:firstMomentsNumberOfCopiesSampleComplexity}
\end{equation}
which is the sample complexity of the estimation of the first moments presented in Theo. \ref{theorem:States}.

\subsubsection{Second-moment estimator}
\label{SectionVAII}

Following the same steps as in the previous subsection, we define the median-of-means estimator $\mathbb{E}_{MM}\left(X\right)$ as
\begin{equation}
\mathbb{E}_{\textup{MM}}\left( X \right) \coloneqq \textup{median}\left\{\mathbb{E}^{(1)}(X), \cdots, \mathbb{E}^{(B)}(X)\right\} \, ,
\end{equation}
with
\begin{equation}
\mathbb{E}^{(\omega)}\left( X \right) \coloneqq \frac{1}{\floor{\mathscr{N}_{2} / B}} \, \sum_{j = (\omega - 1)\floor{\mathscr{N}_{2} / B} + 1}^{\omega \, \floor{\mathscr{N}_{2} / B}} \, X_{\bm{\Sigma}_{j}} \, ,
\end{equation}
where $\bm{\Sigma_{j}} = \{k_{j}, \, l_{j}, \, \Gamma'_{j} \}$, and $\mathscr{N}_{2}$ is the number of measurements required to estimate $\Tr(\mathbf{V}_{t}^{-1} \, \bm{\Gamma}_{p})$ as $\mathbb{E}_{MM}\left(X\right)$ with statistical confidence. Analogously to Eqs. \eqref{eq:probabilityMM} and \eqref{eq:scalingN}, we have
\begin{equation}
\mathscr{N}_{2} = \mathscr{O}\left(\frac{ \mathbb{E}\left(X^{2}\right) \, \ln\left(2 / \Delta \right)}{\epsilon^{2}}\right) \, . \label{eq:secondMomentsNumberOfCopies}
\end{equation}
We leave the demonstration of how to upper-bound $\mathbb{E}\left(X^{2}\right)$ to App. \ref{appendix:StatesSecond}. Here, we present the end result:
\begin{equation}
\mathbb{E}\left(X^{2}\right) \leq 2^{8} \, m^{4} \, s_{t}^{4} \, \Gamma_{\textup{max}}, \label{eq:secondMomentsUpperBoundExpX2}
\end{equation}
where $\Gamma_{\textup{max}}^{2}$ was defined in Sec. \ref{SectionIIIA}. Hence, substituting Eq. \eqref{eq:secondMomentsUpperBoundExpX2} into Eq. \eqref{eq:secondMomentsNumberOfCopies}, we have
\begin{equation}
\mathscr{N}_{2} = \mathscr{O} \left( \frac{m^{4} \, s_{t}^{4} \, \Gamma_{\textup{max}} \, \ln(2 / \Delta)}{\epsilon^{2}} \right), \label{eq:secondMomentsNumberOfCopiesSampleComplexity}
\end{equation}
which is the sample complexity as that of the second moments in Theo. \ref{theorem:States}.

%----------------------------------------------------------------------------------------------
\subsection{Gaussian unitary target channels}
\label{SectionVB}
%----------------------------------------------------------------------------------------------
Similarly to App. \ref{SectionVA}, we can use the framework presented in Eqs. \eqref{eq:expectationMM} - \eqref{eq:scalingN} to upper-bound $\mathscr{N}_{1}^{(c)}$, the number of measurements required to estimate $\mathbb{E}\left(\chi^{(c)}\right)$, as
\begin{equation}
\mathscr{N}_{1}^{(c)} = \mathscr{O}\left(\frac{\mathbb{E}\left(\chi^{(c) \, 2}\right) \, \ln\left(2 / \Delta \right)}{\epsilon^{2}}\right) \, , \label{eq:firstMomentsChannelsNumberOfCopies}
\end{equation}
where estimator $\chi^{(c)}$ was defined in Eq. \eqref{eq:firstMomentsEstimatorChannel}. When we choose the set $\mathscr{S}$ of input states to be composed of $m$-mode coherent states $\{\ket{\bm{\alpha}}\}$, $\mathbb{E}\left(\chi^{(c) \, 2}\right)$ is upper-bounded as
\begin{equation}
\mathbb{E}\left(\chi^{(c) \, 2}\right) \leq 2^{6} \, m^{4} \, s_{\mathscr{U}}^{4} \, E_{max}^{\mathscr{U}} \, E_{max}^{\mathscr{E}} \, , \label{eq:firstMomentsChannelsUpperBoundExpChi2}
\end{equation}
where $s_{\mathscr{U}}$, $E_{max}^{\mathscr{U}}$ and $E_{max}^{\mathscr{E}}$ were defined in Sec. \ref{SectionIIIB}. It is demonstrated in App. \ref{appendix:Channels} that \eqref{eq:firstMomentsChannelsUpperBoundExpChi2} does not depend on the choice of prior probability distribution $\p{\psi} = \{\p{\bm{\alpha}}\}_{\ket{\bm{\alpha}}\in\mathscr{S}}$. Thus, substituting \eqref{eq:firstMomentsChannelsUpperBoundExpChi2} into \eqref{eq:firstMomentsChannelsNumberOfCopies}, we have
\begin{equation}
\mathscr{N}_{1}^{(c)} = \mathscr{O} \left( \frac{m^{4} \, s_{\mathscr{U}}^{4} \, E_{max}^{\mathscr{U}} \, E_{max}^{\mathscr{E}} \, \ln(2 / \Delta)}{\epsilon^{2}} \right) \, , \label{eq:firstMomentsChannelsNumberOfCopiesSampleComplexity}
\end{equation}
which is the sample complexity for first moments presented in Theo. \ref{theorem:Channels}.

As for the second moments, the number of measurements required $\mathscr{N}_{2}^{(c)}$ is such that
\begin{equation}
\mathscr{N}_{2}^{(c)} = \mathscr{O}\left(\frac{\mathbb{E}\left(X^{(c) \, 2}\right) \, \ln\left(2 / \Delta \right)}{\epsilon^{2}}\right) \, , \label{eq:secondMomentsNumberOfCopiesChannels}
\end{equation}
where estimator $X^{(c)}$ was defined in \eqref{eq:secondMomentsChannelsEstimator}. Under the choice of set $\mathscr{S} = \{\ket{\bm{\alpha}}\}$, $\mathbb{E}\left(X^{(c) \, 2}\right)$ can be upper-bounded as
\begin{equation}
\mathbb{E}\left(X^{(c) \, 2}\right) \leq 2^{8} \, m^{4} \, s_{\mathscr{U}}^{4} \, \Gamma_{\textup{max}}^{2} \, , \label{eq:secondMomentsChannelsUpperBoundExpX2}
\end{equation}
where $\Gamma_{\textup{max}}$ was defined in Sec. \ref{SectionIIIB}. As for \eqref{eq:firstMomentsChannelsUpperBoundExpChi2}, App. \ref{appendix:Channels} shows that \eqref{eq:secondMomentsChannelsUpperBoundExpX2} also does not depend on $\p{\psi} = \{\p{\bm{\alpha}}\}_{\ket{\bm{\alpha}}\in\mathscr{S}}$. Thus, substituting Eq. \eqref{eq:secondMomentsChannelsUpperBoundExpX2} into Eq. \eqref{eq:secondMomentsNumberOfCopiesChannels}, we see that
\begin{equation}
\mathscr{N}_{2}^{(c)} = \mathscr{O} \left( \frac{m^{4} \, s_{\mathscr{U}}^{4} \, \Gamma_{\textup{max}} \, \ln(2 / \Delta)}{\epsilon^{2}} \right) \, , \label{eq:secondMomentsChannelsNumberOfCopiesSampleComplexity}
\end{equation}
which, combined with \eqref{eq:firstMomentsChannelsNumberOfCopiesSampleComplexity}, is the sample complexity presented in Theo. \ref{theorem:Channels}.

%----------------------------------------------------------------------------------------------
\subsection{Single-mode applications}
\label{SectionVC}
%----------------------------------------------------------------------------------------------

First, for the coherent-state amplifier, the median-of-means estimator $\mathbb{E}_{\textup{MM}}(\zeta)$ is given by
\begin{equation}
\mathbb{E}_{\textup{MM}}(\zeta) = \textup{median}\left\{ \mathbb{E}^{(1)}(\zeta), \cdots, \mathbb{E}^{(B)}(\zeta) \right\} \, ,
\end{equation}
with
\begin{equation}
\mathbb{E}^{(\omega)}(\zeta) = \frac{1}{\floor{\mathscr{N}_{amp} / B}} \, \sum_{j = (\omega - 1) \floor{\mathscr{N}_{amp} / B} + 1}^{\omega \, \floor{\mathscr{N}_{amp} / B}} \, \zeta_{\bm{\Xi}_{j}}, \label{eq:coherentStateAmplifierEmpiricalMean}
\end{equation}
where $\mathscr{N}_{amp}$ is the number of sampling trials required, and $\bm{\Xi}_{j} = \{ k_{j}, \, \alpha_{j}, \, \nu'_{j} \}\,$. 
Then,
\begin{equation}
\mathscr{N}_{amp} = \mathscr{O}\left(\frac{\mathbb{E}(\zeta^{2}) \, \ln(2 / \Delta)}{\epsilon^{2}}\right). \label{eq:coherentStateAmplifierNumberOfCopies}
\end{equation}
In App. \ref{appendix:CoherentStateAmplifierComplexity} we show that $\mathbb{E}(\zeta^{2})$ is upper-bounded as
\begin{equation}
\mathbb{E}\left( \zeta^{2} \right) \leq \mathscr{S}_{\textup{max}}^{2} \, r_{\textup{max}}^{4} \, , \label{eq:coherentStateAmplifierUpperBoundZeta2}
\end{equation}
with $\mathscr{S}_{\textup{max}}$ and $r_{\textup{max}}^{4}$ defined in Sec. \ref{SectionIIICI}. Hence,
\begin{equation}
\mathscr{N}_{amp} = \mathscr{O}\left( \frac{\mathscr{S}_{\textup{max}}^{2} \, r_{\textup{max}}^{4} \, \ln(2 / \Delta)}{\epsilon^{2}} \right) \, , \label{eq:coherentStateAmplifierNumberOfCopiesSampleComplexity}
\end{equation}
which is the sample complexity obtained in Theo. \ref{theorem:coherentStateAmplifier}.

As for the cubic phase gate, the median-of-means estimator $\mathbb{E}_{\textup{MM}}(Z)$ is
\begin{equation}
\mathbb{E}_{\textup{MM}}(Z) = \textup{median}\left\{ \mathbb{E}^{(1)}(Z), \cdots, \mathbb{E}^{(B)}(Z) \right\} \, ,
\end{equation}
with
\begin{equation}
\mathbb{E}^{(\omega)}(Z) = \frac{1}{\floor{\mathscr{N}_{cub} / B}} \, \sum_{j = (\omega - 1) \floor{\mathscr{N}_{cub} / B} + 1}^{\omega \, \floor{\mathscr{N}_{cub} / B}} \, Z_{\bm{\Phi}_{j}}, \label{eq:cubidPhaseGateEmpiricalMean}
\end{equation}
where $\mathscr{N}_{cub}$ is the number of measurements required and $\bm{\Phi}_{j} = \{ k_{j}, \, \alpha_{j}, \, \mu'_{j} \}$. Moreover,
\begin{equation}
\mathscr{N}_{cub} = \mathscr{O}\left(\frac{\mathbb{E}(Z^{2}) \, \ln(2 / \Delta)}{\epsilon^{2}}\right) \, . \label{eq:cubicPhaseGateNumberOfCopies}
\end{equation}
We let the demonstration of how to upper-bound $\mathbb{E}(Z^{2})$ to App. \ref{appendix:CubicPhaseGateComplexity}. Here, we present the results:
\begin{equation}
\mathbb{E}\left( Z^{2} \right) \leq \mathscr{S}^{'\,2}_{\textup{max}} \, q_{\textup{max}}, \label{eq:cubicPhaseGateUpperBoundZ2}
\end{equation}
where $\mathscr{S}{'}_{\textup{max}}$ and $q_{\textup{max}}$ were defined in Sec. \ref{SectionIIICII}. Therefore,
\begin{equation}
\mathscr{N}_{cub} = \mathscr{O}\left( \frac{\mathscr{S}^{'\,2}_{\textup{max}} \, q_{\textup{max}} \, \ln(2 / \Delta)}{\epsilon^{2}} \right), \label{eq:cubicPhaseGateNumberOfCopiesSampleComplexity}
\end{equation}
which is the sample complexity displayed in Theo. \ref{theorem:cubicPhaseGate}.

%%%%%%%%%%%%%%%%%%%%%%%%%%%%%%%%%%%%%%%%%%%%%%%%%%%%%%%%%%%%%%%%%%%%%%%%%%%%%%%%%%%%%%%%%%%%%%%
\section{Fidelity witness of cubic phase gate}
\label{appendix:FidelityWitnessCubicPhase}
%%%%%%%%%%%%%%%%%%%%%%%%%%%%%%%%%%%%%%%%%%%%%%%%%%%%%%%%%%%%%%%%%%%%%%%%%%%%%%%%%%%%%%%%%%%%%%%

The action of the target gate $\mathscr{U}_{\gamma}$ on a given single-mode coherent state $\ket{\alpha}$ is such that
\begin{eqnarray}
\mathscr{U}_{\gamma}\left(\projector{\alpha}\right) & = & U\left(\gamma\right) \, \projector{\alpha} \, U^{\dagger}\left(\gamma\right) \nonumber \\
& = & U\left(\gamma\right) \, D\left(\alpha\right) \, \projector{0} \, D^{\dagger}\left(\alpha\right) \, U^{\dagger}\left(\gamma\right), \nonumber \\
& & \label{eq:outputCubicPhaseTarget}
\end{eqnarray}
where $U\left(\gamma\right)$, $D\left(\alpha\right)$ and $\ket{0}$ were defined in the main text. The fidelity between $\mathscr{U}_{\gamma}\left(\projector{\alpha}\right)$ and the output of an experimental channel $\mathscr{E}\left(\projector{\alpha}\right)$ is
\begin{eqnarray}
F & \coloneqq & F\left( \mathscr{U}_{\gamma}\left(\projector{\alpha}\right), \, \mathscr{E}\left(\projector{\alpha}\right) \right) \nonumber \\ 
& = & \tr( \mathscr{U}_{\gamma}\left(\projector{\alpha}\right) \, \mathscr{E}\left(\projector{\alpha}\right) ) \nonumber \\
& = & \tr(\projector{0} \, D^{\dagger}\left(\alpha\right) U^{\dagger}\left(\gamma\right) \mathscr{E}\left(\projector{\alpha}\right) U\left(\gamma\right) D\left(\alpha\right)), \nonumber \\
& & \label{eq:fidelityCubicPhaseFirstStep}
\end{eqnarray}
where we have used the cyclicity of the trace operation again. We then have
\begin{equation}
F \geq \mathscr{W}_{\gamma} \coloneqq \tr(W_{\gamma} \, \mathscr{E}\left(\projector{\alpha}\right)) \, , \label{eq:witnessCubicPhaseDefinition}
\end{equation}
where the witness $W_{\gamma}$ was defined in \eqref{eq:witnessCubicPhaseGate}. Moreover, from Baker-Hausdorff Lemma we see that $U\left(\gamma\right) \, a^{\dagger} \, U^{\dagger}\left(\gamma\right) = q - i \, \left( p - 3 \, \gamma \, q^{2} / 2 \right)$. This leads to
\begin{equation}
U\left(\gamma\right) \, a^{\dagger}a \, U^{\dagger}\left(\gamma\right) = q^{2} + \left( p - \frac{3 \, \gamma}{2} \, q^{2} \right)^{2} - \frac{1}{2} \, . \label{eq:relation4}
\end{equation}
Therefore,
\begin{equation}
W_{\gamma} = \frac{3}{2} \, \mathbbm{1}_{2} - \left( p - \frac{3 \, \gamma}{2} \, q^{2} - \Im(\alpha) \right)^{2} - \left(q - \Re(\alpha)\right)^{2} , \label{eq:fidelityCubicPhaseSecondStep}
\end{equation}
where we have used the fact that $\abs{\alpha}^{2} = \Re^{2}(\alpha) + \Im^{2}(\alpha)$. Then, we arrive at \eqref{eq:fidelitySingleCubicPhase} by substituting \eqref{eq:fidelityCubicPhaseSecondStep} into \eqref{eq:witnessCubicPhaseDefinition}.

%%%%%%%%%%%%%%%%%%%%%%%%%%%%%%%%%%%%%%%%%%%%%%%%%%%%%%%%%%%%%%%%%%%%%%%%%%%%%%%%%%%%%%%%%%%%%%%
\section{Scheme for cubic-phase-gate certification}
\label{appendix:CubicPhaseGateSampling}
%%%%%%%%%%%%%%%%%%%%%%%%%%%%%%%%%%%%%%%%%%%%%%%%%%%%%%%%%%%%%%%%%%%%%%%%%%%%%%%%%%%%%%%%%%%%%%%

It is useful to write the observables that appear on the first expectation value on the r.h.s. of \eqref{eq:fidelitySingleCubicPhase} as
\begin{eqnarray}
\left(p - \frac{3 \, \gamma}{2} \, q^{2} \right)^{2} & = & \frac{9 \, \gamma^{2}}{4} \, q^{4} - \frac{3 \, \gamma}{2} \, (q^{2} \, p + p \, q^{2}) + p^{2} \nonumber \\
& = & \frac{9 \, \gamma^{2}}{4} \, q^{4} - 3 \, \gamma \, q \, p q + p^{2} \nonumber \\
& = & \frac{9 \, \gamma^{2}}{4} \, q^{4} + \, \gamma \, p^{3} + p^{2} \nonumber \\
& & \, - \sqrt{2} \, \gamma \, \left[ \left(\frac{q + p}{\sqrt{2}}\right)^{3} - \left(\frac{q - p}{\sqrt{2}}\right)^{3} \right] \, , \label{eq:nonGaussianityRewritten}
\end{eqnarray}
where we have used the relations 
\begin{equation}
q^{2} \, p + p \, q^{2} = 2 \, q \, p \, q = \frac{1}{3} \left( \left(q + p\right)^{3} - \left(q - p\right)^{3} - 2 \, p^{3} \right) \, . \label{eq:relation}
\end{equation}
Thus, substituting \eqref{eq:nonGaussianityRewritten} into \eqref{eq:fidelitySingleCubicPhase}, we have
\begin{widetext}
\begin{eqnarray}
F \geq \mathscr{W}_{\gamma} & = & \frac{3}{2} - \abs{\alpha}^{2} - \frac{9 \, \gamma^{2}}{4} \, \braket{q^{4}}_{\mathscr{E}\left(\projector{\alpha}\right)} + \sqrt{2} \, \gamma \, \left[ \left\langle\left(\frac{q + p}{\sqrt{2}}\right)^{3}\right\rangle_{\mathscr{E}\left(\projector{\alpha}\right)} - \left\langle\left(\frac{q - p}{\sqrt{2}}\right)^{3}\right\rangle_{\mathscr{E}\left(\projector{\alpha}\right)} \right] \nonumber \\
& & - \, \gamma \braket{p^{3}}_{\mathscr{E}\left(\projector{\alpha}\right)} - \, \left( 1 + \, 3 \, \gamma \, \Im(\alpha)\right) \, \braket{q^{2}}_{\mathscr{E}\left(\projector{\alpha}\right)} - \braket{p^{2}}_{\mathscr{E}\left(\projector{\alpha}\right)} \nonumber \\
& & + \, 2 \, \Re(\alpha)\braket{q}_{\mathscr{E}\left(\projector{\alpha}\right)} \, + 2 \, \Im(\alpha) \braket{p}_{\mathscr{E}\left(\projector{\alpha}\right)}, \label{eq:fidelityCubicPhaseGateThirdStep}
\end{eqnarray}
\end{widetext}
where we have used the relation $\left(q - \Re(\alpha)\right)^{2} = q^{2} - 2 \, \Re(\alpha) \, q + \Re^{2}(\alpha)$. As in \eqref{eq:observablesCubicPhase}, if we encode all observables that appear in \eqref{eq:fidelityCubicPhaseGateThirdStep} in a set $\{\mu_{k}\}$ and all their respective coefficients in a set $\{\kappa_{k}\}$, then we arrive at
\begin{equation}
F \geq  \mathscr{W}_{\gamma} = \frac{3}{2} - \abs{\alpha}^{2} - \sum_{k = 1}^{8} \, \kappa_{k} \, \tr( \mu_{k} \, \mathscr{E}\left(\projector{\alpha}\right)) \, . \label{eq:encodedCubicPhaseGate}
\end{equation}
We get the average channel-fidelity witness \eqref{eq:fidelityCubicPhaseGateSampling} by substituting \eqref{eq:encodedCubicPhaseGate} into \eqref{eq:fidelityCubicPhaseAverage}.

%%%%%%%%%%%%%%%%%%%%%%%%%%%%%%%%%%%%%%%%%%%%%%%%%%%%%%%%%%%%%%%%%%%%%%%%%%%%%%%%%%%%%%%%%%%%%%%
\section{Scheme for Gaussian state certification}
\label{appendix:secondMomentsMeasurementScheme}
%%%%%%%%%%%%%%%%%%%%%%%%%%%%%%%%%%%%%%%%%%%%%%%%%%%%%%%%%%%%%%%%%%%%%%%%%%%%%%%%%%%%%%%%%%%%%%%

Following the definition of $\bm{\Gamma}_{p}$ given in \eqref{eq:secondMoments}, it is indispensable to separate the elements $\Gamma_{kl}$ into two categories \cite{Aspuru2012, Aolita2015}:
\begin{enumerate}
\item $(k, \, l) \neq (2j-1, 2j), \forall \, j \in [m]$: Single-mode observables $q_{k}^{2}$ and $p_{k}^{2}$, plus two-body observables $q_{k} \, q_{l}$ , $q_{k} \, p_{l}$ , and $p_{k} \, p_{l}$ that can be measured by simultaneously homodyning modes $k$ and $l$ independently. Here, $(\bm{\Gamma}_{p})_{kl}$ can be expressed as
\begin{equation}
\left( \bm{\Gamma}_{p} \right)_{kl} = \int \, d\Gamma' \, \p{\Gamma' \, | \, k, \, l} \, {\Gamma}_{kl}' \, ,
\end{equation}
where $\Gamma_{kl}' = \left(r_{k}'r_{l}' + r_{l}' r_{k}'\right) / 2$ is the possible measured eigenvalue, and $\p{\Gamma' \, | \, k, \, l} = \tr(\Pi_{k,l,\Gamma'} \, \varrho_{p})$.

\item $(k, \, l) = (2j-1, 2j), \forall \, j \in [m]$: Single-mode observables of the form $\left(q_{j} \, p_{j} + p_{j} \, q_{j}\right) / 2$. As single-mode field quadratures $q_{j}$ and $p_{j}$ do not commute, it is necessary to indirectly estimate this kind of bilinear observable by using the relation 
\begin{equation}
\quad \quad \frac{1}{2} \left( q_{j} \, p_{j} + p_{j} \, q_{j} \right) = \left( \frac{q_{j} + p_{j}}{\sqrt{2}} \right)^{2} - \frac{1}{2} \left( q_{j}^{2} + p_{j}^{2} \right) \, . \label{eq:secondMomentsDecomposition}
\end{equation}
To estimate the first term on the r.h.s of \eqref{eq:secondMomentsDecomposition}, a certifier can homodyne, in a single setting, each mode $j$ independently in a rotated quadrature $(q_{j} + p_{j}) / \sqrt{2}$, while $q_{j}^{2}$ and $p_{j}^{2}$ can be estimated as mentioned in the category above. Thus, in this case, we can write
\begin{equation}
\quad \quad \left( \bm{\Gamma}_{p} \right)_{2j-1, 2j} = \sum_{y = 1}^{3} \, \int \, d\eta_{y}' \, \p{\eta_{y}' \, | \, 2\,j-1, \, 2\,j } \, \eta_{y}' \, ,
\end{equation}
where $\,\eta_{y}'$ is the eigenvalue of observable $\eta_{y} \in \{ (q_{j} + p_{j})/\sqrt{2}, \, q_{j}, \, p_{j} \}\,$, and $\p{\eta_{y}' \, | \, 2\,j-1, \, 2\,j } = \tr(\varrho_{p} \, \Pi_{2j-1, \, 2j, \, \eta_{y}'})$.
\end{enumerate}

%%%%%%%%%%%%%%%%%%%%%%%%%%%%%%%%%%%%%%%%%%%%%%%%%%%%%%%%%%%%%%%%%%%%%%%%%%%%%%%%%%%%%%%%%%%%%%%
\section{Upper bounds in Gaussian-state witnesses}
\label{appendix:States}
%%%%%%%%%%%%%%%%%%%%%%%%%%%%%%%%%%%%%%%%%%%%%%%%%%%%%%%%%%%%%%%%%%%%%%%%%%%%%%%%%%%%%%%%%%%%%%%

%----------------------------------------------------------------------------------------------
\subsection{First Moments}
\label{appendix:StatesFirst}
%----------------------------------------------------------------------------------------------

Here we show that $\mathbb{E}\left(\chi^{2}\right)$ can be expressed as the product of the Frobenius norm of $\mathbf{V}_{t}^{-1}$ as well as the Euclidean norm of $\mathbf{x}_{t}$ and the trace of $\bm{\Gamma}_{p}$:
\begin{eqnarray}
\mathbb{E}\left(\chi^{2}\right) & = & \sum_{k,l} \, \int \, dr' \, \p{k, \, l, \, r'} \, \chi_{k, l, r'}^{2} \nonumber \\
& = & \sum_{k,l} \, \int \, dr' \, \p{k,\,l,\,r'} \, \frac{{r_{k}'}^{2} \, \left(\mathbf{x}^{T}\right)_{l}^{2}}{\left(\mathbf{V}_{t}^{-1}\right)_{kl}^{2}} \, \left|\left|\mathbf{V}_{t}^{-1}\right|\right|_{F}^{4} \nonumber \\
& = & \left|\left|\mathbf{V}_{t}^{-1}\right|\right|_{F}^{2} \, \sum_{k,l} \, \int \, dr' \, \p{r' \, | \, k, \, l} \, {r_{k}'}^{2} \, \left(\mathbf{x}^{T}\right)_{l}^{2} \nonumber \\
& = & \left|\left|\mathbf{V}_{t}^{-1}\right|\right|_{F}^{2} \, \left|\left| \mathbf{x}_{t} \right|\right|_{2}^{2} \, \sum_{k} \, \int \, dr' \, \p{r' \, | \, k} \, {r_{k}'}^{2} \nonumber \\
& = & \Tr(\bm{\Gamma}_{p}) \, \left|\left| \mathbf{x}_{t} \right|\right|_{2}^{2} \, \left|\left|\mathbf{V}_{t}^{-1}\right|\right|_{F}^{2}. \label{eq:meanChi2}
\end{eqnarray}
From \eqref{eq:secondMoments}, we can write $\Tr(\bm{\Gamma}_{p})$ as
\begin{equation}
\Tr(\bm{\Gamma}_{p}) = \sum_{k = 1}^{m} \, \left( \braket{q_{k}^{2}}_{\varrho_{p}} + \braket{p_{k}^{2}}_{\varrho_{p}} \right) \leq m \, E_{max}^{(p)}, \label{eq:upperBoundGammaPreparation}
\end{equation}
where $E_{max}^{(p)}$ is maximum single-mode energy of a preparation state among all $m$ modes. Moreover, the combination of \eqref{eq:WilliamsonEuler} and \eqref{eq:SingleModeSqueezers} with the Cauchy-Schwarz inequality \cite{Steele2004} leads to
\begin{eqnarray}
\left|\left|\mathbf{V}_{t}^{-1}\right|\right|_{F}^{2} & = & \Tr(\mathbf{V}_{t}^{-2}) \nonumber \\
& \leq & \left[\Tr(\mathbf{V}_{t}^{-1})\right]^{2} = \left[ 8 \, \sum_{k=1}^{m} \, \cosh\left(2 \, \xi_{k}^{(t)} \right) \right]^{2} \nonumber \\
& \leq & 2^{6} \, m^{2} \, \cosh^{2}\left(2 \, \xi_{max}^{(t)} \right) \nonumber \\
& \leq & 2^{6} \, m^{2} \, s_{t}^{4} \, , \label{eq:upperBoundCovarianceTarget}
\end{eqnarray}
where $s_{t} \coloneqq \exp\left(\xi_{max}^{(t)}\right)$, and with $\xi_{max}^{(t)} = \max_{k} \, \xi_{k}^{(t)}$ being the maximum single-mode squeezing parameter among all $m$ modes of a target state $\varrho_{t}$. Therefore, substituting \eqref{eq:upperBoundGammaPreparation} and \eqref{eq:upperBoundCovarianceTarget} into \eqref{eq:meanChi2}, we arrive at \eqref{eq:firstMomentsUpperBoundExpChi2}.

%----------------------------------------------------------------------------------------------
\subsection{Second Moments}
\label{appendix:StatesSecond}
%----------------------------------------------------------------------------------------------

The demonstration of the upper bound of $\mathbb{E}\left( X^{2} \right)$ is straightforward:
\begin{eqnarray}
\mathbb{E}\left( X^{2} \right) & = & \sum_{k,\,l} \, \int \, d\Gamma' \, \p{k, \, l, \, \Gamma'} \, X_{k, \, l, \, \Gamma'}^{2} \nonumber \\
& = & \sum_{k,\,l} \, \int \, d\Gamma' \, \p{k, \, l, \, \Gamma'} \, \frac{\Gamma_{kl}'^{\,2}}{(\mathbf{V}_{t}^{-1})_{kl}^{2}} \, \left|\left|\mathbf{V}_{t}^{-1}\right|\right|_{F}^{4} \nonumber \\
& = & \left|\left|\mathbf{V}_{t}^{-1}\right|\right|_{F}^{2} \, \sum_{k,\,l} \, \int \, d\Gamma' \, \p{\Gamma' \, | \, k, \, l} \, \Gamma_{kl}'^{\,2} \nonumber \\
& = & \left|\left|\mathbf{V}_{t}^{-1}\right|\right|_{F}^{2} \, \sum_{k,\,l} \, \tr(\Gamma_{kl}^{\,2} \, \varrho_{p}) \nonumber \\
& \leq & 4 \, m^{2} \, \Gamma_{\textup{max}} \, \left|\left|\mathbf{V}_{t}^{-1}\right|\right|_{F}^{2} \nonumber \\
& \leq & 2^{8} \, m^{4} \, s_{t}^{4} \, \Gamma_{\textup{max}} \, , \label{eq:upperBoundMeanX2}
\end{eqnarray}
where $\Gamma_{\textup{max}} \coloneqq \max_{kl} \, \tr(\Gamma_{kl}^{2} \, \varrho_{p})$ was defined in the main text, and we have used \eqref{eq:upperBoundCovarianceTarget} in the last step.

%%%%%%%%%%%%%%%%%%%%%%%%%%%%%%%%%%%%%%%%%%%%%%%%%%%%%%%%%%%%%%%%%%%%%%%%%%%%%%%%%%%%%%%%%%%%%%%
\section{Upper bounds in Gaussian-channel witnesses}
\label{appendix:Channels}
%%%%%%%%%%%%%%%%%%%%%%%%%%%%%%%%%%%%%%%%%%%%%%%%%%%%%%%%%%%%%%%%%%%%%%%%%%%%%%%%%%%%%%%%%%%%%%%

Analogously to App. \ref{appendix:States}, we start upper-bounding $\mathbb{E}\left( \chi^{(c) 2} \right)$ from
\begin{equation}
\mathbb{E}\left( \chi^{(c) \, 2} \right) = \sum_{\ket{\bm{\alpha}}\in\mathscr{S}} \p{\bm{\alpha}} \, \Tr(\bm{\Gamma}_{\mathscr{E}}) \, \left|\left| \mathbf{x}_{\mathscr{U}} \right|\right|_{2}^{2} \, \left|\left| \mathbf{V}_{\mathscr{U}}^{-1} \right|\right|_{F}^{2}. \label{eq:upperBoundMeanChi2ChannelsFirstStep}
\end{equation}
Respectively, $\Tr(\bm{\Gamma}_{\mathscr{E}})$ and $\left|\left| \mathbf{V}_{\mathscr{U}}^{-1} \right|\right|_{F}^{2}$ are upper-bounded by \eqref{eq:upperBoundGammaPreparation} and \eqref{eq:upperBoundCovarianceTarget}. The squared norm $\left|\left| \mathbf{x}_{\mathscr{U}} \right|\right|_{2}^{2}$ can be upper-bounded as
\begin{eqnarray}
\left|\left| \mathbf{x}_{\mathscr{U}} \right|\right|_{2}^{2} & = & \sum_{k=1}^{m} \left( \braket{q_{k}^{2}}_{\mathscr{U}(\alpha)} + \braket{p_{k}^{2}}_{\mathscr{U}(\alpha)}\right) \nonumber \\
& \leq & \sum_{k=1}^{m} \, E_{k,\,\textup{max}}^{\mathscr{U}}\left(\ket{\bm{\alpha}}\right) \nonumber \\
& \leq & m \, E_{\textup{max}}^{\mathscr{U}}, \label{eq:upperBoundFirstMomentsNormChannels}
\end{eqnarray}
where $E_{k,max}^{\mathscr{U}}\left(\ket{\bm{\alpha}}\right)$ is the $k$-th maximum single-mode energy as a function of the input states $\ket{\bm{\alpha}}$, and 
\begin{equation}
E_{max}^{\mathscr{U}} \coloneqq \max_{k, \, \ket{\bm{\alpha}} \in \mathscr{S}} \, E_{k, \, \textup{max}}^{\mathscr{U}}\left(\ket{\bm{\alpha}}\right) \label{eq:maximumEnergy}
\end{equation}
is the maximum single-mode energy among all $m$ modes and input states. Substituting \eqref{eq:upperBoundGammaPreparation}, \eqref{eq:upperBoundCovarianceTarget} and \eqref{eq:upperBoundFirstMomentsNormChannels} into \eqref{eq:upperBoundMeanChi2ChannelsFirstStep}, and considering normalized distributions, we arrive at \eqref{eq:firstMomentsChannelsUpperBoundExpChi2}.

Regarding $\mathbb{E}(X^{(c) \, 2})$, we use \eqref{eq:upperBoundMeanX2} to start from
\begin{equation}
\mathbb{E}\left( X^{(c) \, 2} \right) = \mathop{\sum_{\ket{\bm{\alpha}}\in\mathscr{S}}}_{k,\,l} \, \p{\bm{\alpha}} \, \left|\left|\mathbf{V}_{\mathscr{U}}^{-1} \right|\right|_{F}^{2} \, \tr(\Gamma_{kl}^{2} \, \mathscr{E}\left(\projector{\bm{\alpha}}\right)) \, . \label{eq:upperBoundMeanX2Channels}
\end{equation}
Consider that
\begin{equation}
\sum_{k, \, l} \, \tr(\Gamma_{kl}^{2} \, \mathscr{E}\left(\projector{\bm{\alpha}}\right)) \leq 4 \, m^{2} \, \Gamma_{\textup{max}} \, , \label{eq:upperBoundGammaSquared}
\end{equation}
with $\Gamma_{\textup{max}} \coloneqq \max_{k,\,l,\,\ket{\bm{\alpha}} \in \mathscr{S}}\,\tr(\Gamma_{kl}^{2} \, \mathscr{E}\left(\projector{\bm{\alpha}}\right))$ being the maximum expectation value w.r.t. $\mathscr{E}\left(\projector{\bm{\alpha}}\right)$, among all $(k, \, l)$ and $\projector{\bm{\alpha}}\,$. Then, substituting \eqref{eq:upperBoundCovarianceTarget} and \eqref{eq:upperBoundGammaSquared} into \eqref{eq:upperBoundMeanX2Channels}, we get \eqref{eq:secondMomentsChannelsNumberOfCopiesSampleComplexity}.

%%%%%%%%%%%%%%%%%%%%%%%%%%%%%%%%%%%%%%%%%%%%%%%%%%%%%%%%%%%%%%%%%%%%%%%%%%%%%%%%%%%%%%%%%%%%%%%
\section{Upper bounds in coherent-state-amplifier witness}
\label{appendix:CoherentStateAmplifierComplexity}
%%%%%%%%%%%%%%%%%%%%%%%%%%%%%%%%%%%%%%%%%%%%%%%%%%%%%%%%%%%%%%%%%%%%%%%%%%%%%%%%%%%%%%%%%%%%%%%

From \eqref{eq:observablesCoherentStateAmplifier} to \eqref{eq:ExpectationValueZeta}, we can write $\mathbb{E}(\zeta^{2})$ as
\begin{eqnarray}
\mathbb{E}\left( \zeta^{2} \right) & = & \sum_{k,\,\mathscr{S}} \, \int d\nu' \, \p{k, \, \alpha, \, \nu'} \, {\nu'}^{\, 2} \, \left[ \sum_{l} \abs{\tau_{l}} \right]^{2} \nonumber \\
& = & \sum_{k,\,\mathscr{S}} \, \p{\alpha} \, \abs{\tau_{k}} \, \braket{\nu_{k}^{2}}_{\mathscr{E}\left(\projector{\alpha}\right)} \, \left[ \sum_{l} \abs{\tau_{n}} \right] \label{eq:upperBoundMeanZeta2FirstStep}
\end{eqnarray}
We can upper-bound $\braket{\nu_{k}^{2}}_{\mathscr{E}\left(\projector{\alpha}\right)}$ as
\begin{equation}
\braket{\nu_{k}^{2}}_{\mathscr{E}\left(\projector{\alpha}\right)} \leq r_{\textup{max}}^{4} \coloneqq \max_{k, \, \ket{\alpha} \in \mathscr{S}} \tr(\nu_{k}^{2} \, \mathscr{E}\left(\projector{\alpha}\right)). \label{eq:coherentStateAmplifierUpperBoundObservables}
\end{equation}
We can also write the bound
\begin{equation}
\sum_{l} \, \abs{\tau_{l}} \leq \mathscr{S}_{\textup{max}} \coloneqq 2 \, \left( 1 + \max_{\ket{\alpha} \in \mathscr{S}} \abs{\Re(\alpha)} + \max_{\ket{\alpha} \in \mathscr{S}} \abs{\Im(\alpha)} \right) , \label{eq:coherentStateAmplifierUpperBoundCoefficients}
\end{equation}
then it is straightforward to see that
\begin{equation}
\mathbb{E}\left( \zeta^{2} \right) \leq \mathscr{S}_{\textup{max}}^{2} \, r_{\textup{max}}^{4}. \label{upperBoundMeanZeta2}
\end{equation}

%%%%%%%%%%%%%%%%%%%%%%%%%%%%%%%%%%%%%%%%%%%%%%%%%%%%%%%%%%%%%%%%%%%%%%%%%%%%%%%%%%%%%%%%%%%%%%%
\section{Upper bounds in cubic-phase-gate witness}
\label{appendix:CubicPhaseGateComplexity}
%%%%%%%%%%%%%%%%%%%%%%%%%%%%%%%%%%%%%%%%%%%%%%%%%%%%%%%%%%%%%%%%%%%%%%%%%%%%%%%%%%%%%%%%%%%%%%%

From \eqref{eq:observablesCubicPhase} to \eqref{eq:ExpectationValueZ}, we can write $\mathbb{E}(Z^{2})$ as
\begin{eqnarray}
\mathbb{E}\left( Z^{2} \right) & = & \sum_{k,\,\mathscr{S}} \, \int d\mu' \, \p{k, \alpha, \, \mu'} \, {\mu'}^{\, 2} \, \left[ \sum_{l} \abs{\kappa_{l}} \right]^{2} \nonumber \\
& = & \sum_{k,\,\mathscr{S}} \, \p{k, \, \alpha} \, \left[ \sum_{l} \abs{\kappa_{l}} \right]^{2} \, \braket{\mu_{k}^{2}}_{\mathscr{E}\left(\projector{\alpha}\right)} \, \label{eq:upperBoundMeanZ2FirstStep}
\end{eqnarray}
Again, we can upper-bound $\braket{\mu_{k}^{2}}_{\mathscr{E}\left(\projector{\alpha}\right)}$ as
\begin{equation}
\braket{\mu_{k}^{2}}_{\mathscr{E}\left(\projector{\alpha}\right)} \leq q_{\textup{max}} \coloneqq \max_{k, \, \ket{\alpha} \in \mathscr{S}} \tr(q^{8} \, \mathscr{E}\left(\projector{\alpha}\right)). \label{eq:cubicPhaseGateUpperBoundObservables}
\end{equation}
We can also see that 
\begin{equation}
\label{eq:cubicPhaseGateUpperBoundCoefficients}
\sum_{l} \abs{\kappa_{l}} \leq \mathscr{S}{'}_{\textup{max}}
\end{equation}
where $\mathscr{S}{'}_{\textup{max}}$ was defined in \eqref{eq:boundSet2}.
Then, substituting \eqref{eq:cubicPhaseGateUpperBoundObservables} and \eqref{eq:cubicPhaseGateUpperBoundCoefficients} into \eqref{eq:upperBoundMeanZ2FirstStep}, we have
\begin{equation}
\mathbb{E}\left( Z^{2} \right) \leq \mathscr{S}{'}_{\textup{max}}^{2} \, q_{\textup{max}}. \label{upperBoundMeanZ2}
\end{equation}

\end{document}